\documentclass[apj,numberedappendix]{emulateapj}
\usepackage{amssymb,amsmath}
\usepackage{graphicx}
\usepackage{appendix}
\usepackage{natbib}
\usepackage{hyperref}
\usepackage{enumerate}
\usepackage{multirow}
\usepackage{bm}
\usepackage{url}

\begin{document}


\title{Extended, Dusty Star formation Fueled by a Residual Cooling Flow \\in the Cluster of Galaxies S\'ersic~159-03}

%\author{The SPT TeamS\'ersic}
%auto-ignore

\author{M.\ McDonald$^{1}$, N.\ Werner$^{2}$, J.B.R.\ Oonk$^{3}$, and S.\ Veilleux$^{4,5}$}
\affil{$^1$MIT Kavli Institute for Astrophysics and Space Research, MIT, 77 Massachusetts Avenue, Cambridge, MA 02139, USA \email{Email: mcdonald@space.mit.edu}   \\
$^2$Kavli Institute for Particle Astrophysics and Cosmology, Stanford University, 452 Lomita Mall, Stanford, CA 94305-4085, USA\\
$^3$ASTRON, Netherlands Institute for Radio Astronomy, PO Box 2, NL-7990 AA Dwingeloo, the Netherlands\\
$^4$Department of Astronomy, University of Maryland, College Park, MD 20742, USA\\
$^5$Joint Space-Science Institute, University of Maryland, College Park, MD 20742, USA
}

\email{Email: mcdonald@space.mit.edu}


\begin{abstract}
While the cooling of the hot intra-cluster medium (ICM) in the cores of galaxy clusters is mostly counteracted by heating from the central active galactic nucleus (AGN), the balance is not perfect. This can lead to residual cooling flows and low-level star formation, the physics of which is not well understood. 
Here we present a detailed study of the residual cooling flow in the center of the low mass galaxy cluster S\'ersic 159-03 (A~S1101) using far-ultraviolet imaging from the \emph{Hubble Space Telescope} and far-infrared (FIR) spectroscopy and photometry from the \emph{Herschel} space observatory, along with a wealth of archival data. We detect extended emission at UV, FIR, and [C\,\textsc{ii}], indicating a star formation rate of $\sim$1--3 $M_\odot$~yr$^{-1}$, depending on the indicator and assumptions made. The most recently formed stars appear spatially coincident with the lowest entropy ICM. We speculate that this low-entropy gas has been displaced by the central AGN $\sim$7.5~kpc north of the cD galaxy. These data demonstrate that the displacement of the cooling core from the direct vicinity of the central AGN can temporarily break the feedback cycle and lead to cooling and star formation that is offset from the center of the galaxy. 
We find an abundance ($\sim$10$^7$~M$_\odot$) of cold (20\,K) dust in the center of the cluster and a second FIR peak $\sim$30\,kpc to the north of the central galaxy.
If confirmed to be associated with the cooling filaments, this would be the most extended complex of dust yet found in a cool core cluster.
%. Additionally, extended FIR continuum emission is observed north to north-east of the central galaxy. If this FIR emission is located at the distance of the cluster then it implies that large amounts ($\sim$10$^8$ M$_\odot$) of cold (12\,K) dust may be associated with the H$\alpha$ filament $\sim$30 kpc north of the central galaxy. 

\end{abstract}


%================================================================%
%============== INTRODUCTION ====================================%
%================================================================%
\section{Introduction}
\setcounter{footnote}{0}
Groups and clusters of galaxies, with central cooling times of the hot X-ray emitting plasma shorter than the Hubble time, often host central dominant galaxies surrounded by bright, extended, filamentary optical emission-line nebulae \citep[e.g.,][]{johnstone87,crawford99,edwards07,mcdonald10}. These filaments of ionized gas are usually associated with both warm (1000--2000~K) molecular hydrogen seen in the near-infrared \citep[e.g.][]{jaffe97,falcke98,donahue00,edge02,hatch05,jaffe05,johnstone07,oonk10,lim12} and with cold (30--100~K) atomic and molecular gas traced by [\ion{C}{2}] \citep{edge10a,mittal11,mittal12,werner14} and CO \citep[e.g.][]{edge01,edge03,salome03,mcdonald12b}. This cold gas is most likely produced by thermally unstable cooling from the hot phase \citep[e.g.][]{voit08,gaspari12,sharma12,werner14}.

Many of these systems are experiencing star formation at rates of a few M$_\odot$~year$^{-1}$ \citep[e.g.][]{hicks05,odea08, mcdonald11b}. In rare, extreme cases, star formation can reach rates of up to 800~M$_\odot$~year$^{-1}$ \citep{mcdonald12c}. However, the global star-forming efficiency of the cold gas clouds at the centers of cool core clusters is relatively low, of order $\sim$10\% \citep{mcdonald11b,mcdonald14}. Star-forming brightest cluster galaxies are also often experiencing powerful outbursts of their central active galactic nuclei \citep[AGN; e.g.,][]{mcnamara00,mcnamara09,ehlert11}. The detailed knowledge of the mechanisms that connect cooling, AGN feedback, and star formation is key for our understanding of the formation and evolution of early type galaxies.

Here we explore the relation between cooling, AGN feedback, and star formation in the low mass cooling-core cluster of galaxies S\'ersic 159-03 (A~S1101). The thermal properties and chemical composition of its hot intra-cluster medium (ICM) have been studied in detail using all current, major X-ray observatories: \emph{XMM-Newton} \citep{kaastra2001,deplaa2006}, \emph{Suzaku} \citep{werner07} and \emph{Chandra} \citep{werner11}. While the large scale X-ray morphology of the cluster is relatively relaxed, its core is strongly disturbed. The cluster center displays signs of powerful AGN feedback, which has mostly cleared the central $r < 7.5$~kpc of the dense, X-ray emitting ICM \citep{werner11}. The relatively modest (factor of 1.5) temperature drop in the core of the cluster also indicates exceptionally powerful AGN activity in this system \citep{sun09a}. The central dominant galaxy of the cluster shows dust lanes, molecular gas, excess UV emission, and a bright, 44~kpc long H$\alpha$+[\ion{N}{2}] filamentary nebula, associated with low entropy, high metallicity, cooling X-ray emitting gas \citep{werner11,oonk10,mcdonald12a,farage12}. \citet{werner11} propose that the low entropy ICM has been uplifted by the powerful `radio mode' AGN, and subsequently, because it has been removed from the direct influence of the AGN jets, it cools, falls back and forms stars.

In this paper, we study the cooling and star formation in S\'ersic~159-03 using new far-ultra violet (FUV) images from the \emph{Hubble Space Telescope} (\emph{HST}) and far-infrared (FIR) data from \emph{Herschel}, along with a wealth of archival data. Throughout the paper we assume $z=0.058$ for S\'ersic~159-03, and $\Lambda$CDM cosmology with H$_0$ = 73 km s$^{-1}$ Mpc$^{-1}$, $\Omega_M$ = 0.27, and $\Omega_{\Lambda}$ = 0.73.
\section{Data}

\begin{figure*}
\centering
\includegraphics[width=0.99\textwidth]{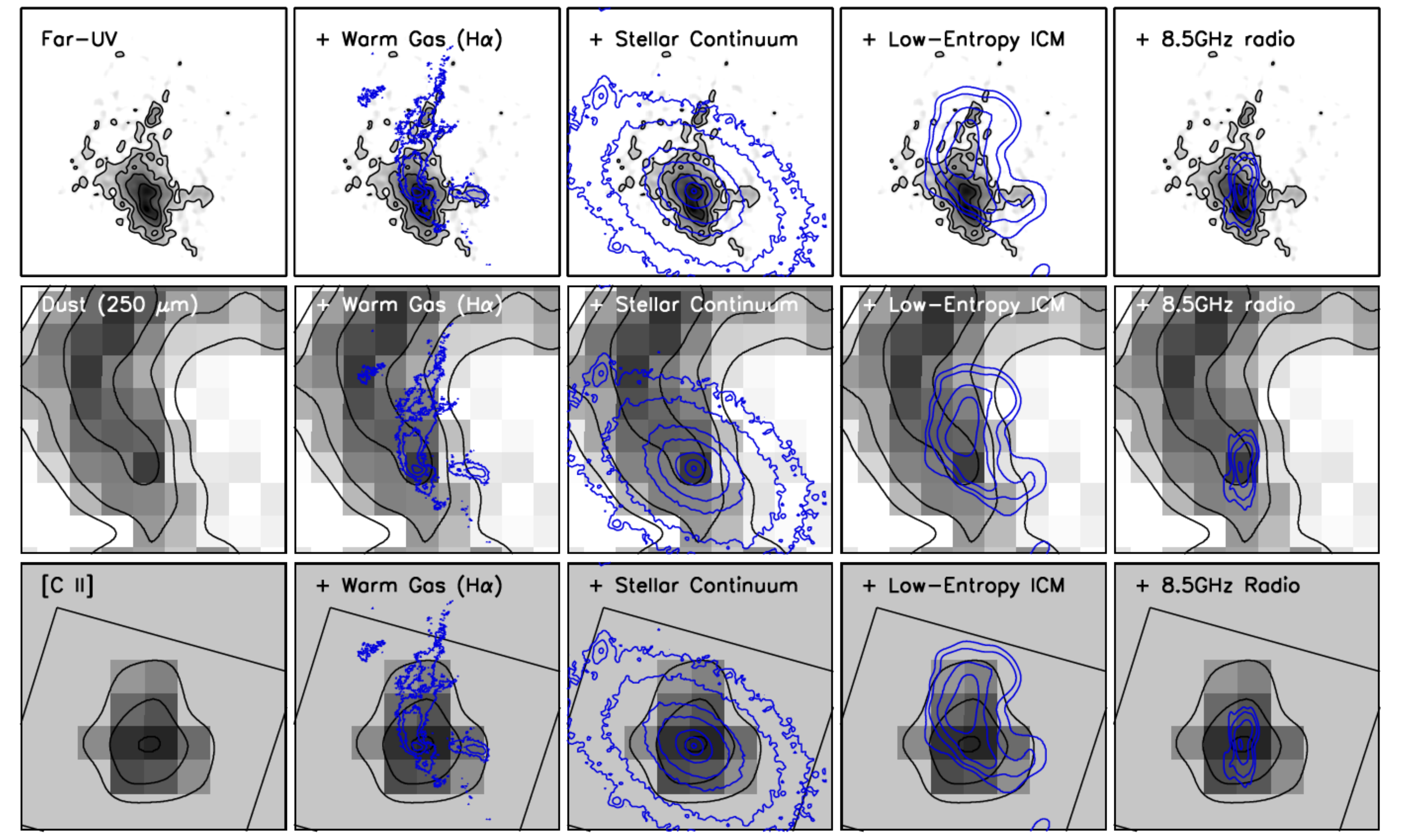}
\caption{Far-UV (F140LP-F150LP), Far-IR (250$\mu$m), and [C\,\textsc{ii}] images of the central galaxy in S\'ersic~159-03. All panels are 50$^{\prime\prime}$ (53\,kpc) on a side. In the left-most column we show the images with overlaid contours to highlight low surface brightness features. In subsequent columns, we overlay blue contours representing emission from H$\alpha$, old stars, low-entropy intracluster plasma, and radio jets. In the lower row, we outline the field of view of the PACS spectroscopy.}
\label{fig:imtile}
\end{figure*}

In this paper, we utilize data across a wide range of wavelengths in an effort to explain the multiphase gas in the core of S\'ersic~159-03. Below, we describe the acquisition and reduction of new data from the \emph{Herschel} and \emph{Hubble} space telescopes, along with supporting data at X-ray, optical, and radio wavelengths published in previous works.

\subsection{HST ACS/SBC Far UV Imaging}
\begin{figure}
\centering
\vspace{0.9cm}
%\begin{minipage}{0.49\textwidth}
\includegraphics[width=0.45\textwidth,trim=1cm 9.5cm 2cm 3cm]{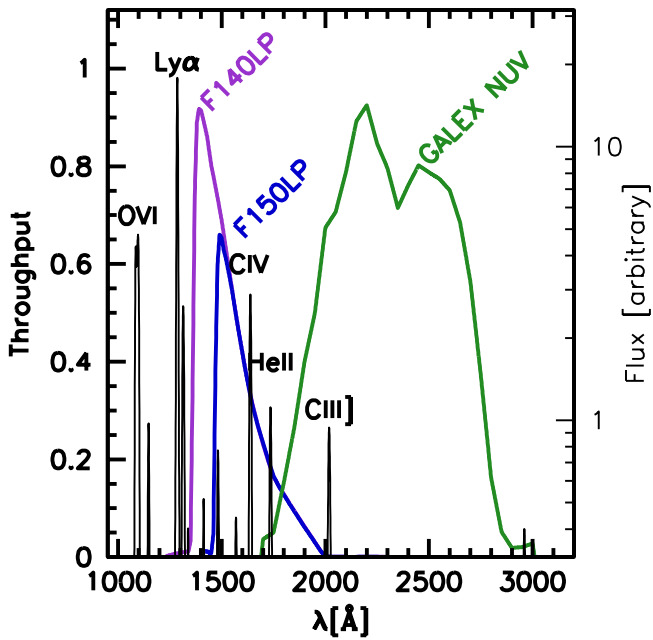}
%\end{minipage}
\vspace{2cm}
\caption{Filter throughput curves for two far-UV filters on HST ACS/SBC and the GALEX near-UV filter. In black, we show the UV spectrum of NGC~1068, a nearby AGN. This figure demonstrates that the near-UV filter is relatively free of line emission, as is the difference of the two ACS/SBC long-pass filters.}
\label{fig:uvfiltesr}
\end{figure}

Far-UV imaging was acquired using the Advanced Camera for Surveys Solar Blind Channel (herafter ACS/SBC) on the \emph{Hubble Space Telescope} (hereafter HST) in both the F140LP and F150LP bandpasses (PID \#12570; PI Veilleux). The total exposure time in each filter was 1302s. Both pointings were centered at ($\alpha,\delta$) = 348.49384$^{\circ}$, -42.7251$^{\circ}$, which provided overlap with the full extent of the complex H$\alpha$ emission presented in \cite{mcdonald10}. The choice of two far-UV filters was made for two reasons. First, the use of two filters allows us to remove the known ACS/SBC red leak, which has a non-negligible contribution due to the fact that the underlying BCG is very red and luminous. Since both F140LP and F150LP are long-pass filters, they have approximately the same throughput at $\gtrsim$1500\AA. Thus, the difference of these two filters yields a narrow effective bandpass from 1400--1500\AA, which is free from any red leak emission. Secondly, as we show in Figure \ref{fig:uvfiltesr}, this narrow bandpass is free of contamination from bright UV emission lines such as Ly$\alpha$, C\,\textsc{iv}, and He\,\textsc{ii}. Thus, we can conclude that any emission in the differenced image is most likely due to UV continuum.

Following the differencing of the F140LP and F150LP images, we bin the images $8\times8$ and smooth with a 1.5 pixel smoothing radius, yielding matching angular resolution at far-UV and H$\alpha$ \citep[see also][]{mcdonald11b}. This process improves the relatively poor quality of the raw data, allowing the detection of extended, filamentary features in the binned data. The final processed (differenced, binned, smoothed) image is shown in Figure \ref{fig:imtile}, compared to imaging at a variety of other wavelengths.

\subsection{Herschel PACS \& SPIRE Imaging}
We observed the FIR continuum emission in S\'ersic~159-03 with the Photodetector 
Array Camera and Spectrometer \citep[PACS;][]{poglitsch10} and the 
Spectral and Photometric Imaging Receiver \citep[SPIRE;][]{griffin10}.

PACS imaging observations (OBSIDs: 1342232470, 1342232471, 1342232472, 1342232473) were conducted in large-scan
mapping mode with a medium speed of 20 arcsec/s in the blue-short (BS;
70~$\mu$m), blue-long (BL; 100~$\mu$m) and the red (R; 160~$\mu$m) 
filters. At these wavelengths a resolution of about 5, 8 and 12 
arcsec, respectively, is obtained. Each of the scans consisted of 10  
scan line legs of 2 arcmin length with a cross-scan step of 4 arcsec. 
The scan and orthogonal cross-scan observations were reduced using the 
HIPE software version 8.2.0, using the PACS pipeline scanmap script for 
extended emission. Default settings for high-pass filter and the 
photometric projection task (PhotProject) were used. The script processes 
the data from level~0 (raw data) to level~2 (flux calibrated maps). The 
reduced and calibrated PACS maps have a size 3.5 arcmin by 7.3 arcmin. The 
uncertainty on the absolute flux scale is estimated to be less than 5 
percent in all three PACS bands \citep{balog14,muller14}.

The SPIRE imaging observations (OBSID: 1342232366) were conducted in small-scan
mapping mode. Simultaneous data at 250, 350 and 500\,$\mu$m was taken. At 
these wavelengths a resolution of about 18, 25 and 36 arcsec, respectively, 
is obtained. The observations were reduced using the HIPE software version 
8.2.0, using the SPIRE POF10 pipeline script with default settings. This script 
processes the data from level~0 (raw data) to level~2 (flux calibrated maps).
The reduced and calibrated SPIRE maps have a size 9.4 arcmin by 12.2 arcmin.
The uncertainty on the absolute flux scale is estimated to be less than 5 
percent in all three SPIRE bands \citep{bendo13}.

\subsection{Herschel PACS Spectroscopy}
We observed the FIR [C~II]$\lambda$157$\mu$m cooling line in S\'ersic~159-03
with the PACS integral-field spectrometer \citep{poglitsch10}
on the Herschel Space Observatory (OBSID: 1342234062). The observations were taken
in line spectroscopy mode with chopping-nodding to remove
the telescope background, sky background and dark current.
A chopper throw of 6 arcmin was used. The observations were
taken in pointed mode, targeting the center of the CC BCG.
The observations were reduced using the HIPE software
version 8.2.0, using the PACS ChopNodLineScan pipeline
script. This script processes the data from level~0 (raw channel
data) to level~2 (flux calibrated spectral cubes). The 
uncertainty on the absolute flux scale for PACS spectroscopy is 
estimated to be less than 30 percent\footnote{\url{http://herschel.esac.esa.int/twiki/pub/Public/PacsCalib}\\\url{rationWeb/PacsSpectroscopyPerformanceAndCalibration_v2_4.pdf}}.

During the final stage of the reduction, the data was spectrally 
and spatially rebinned into a 5~$\times$~5~$\times$~$\lambda$ cube. 
This we will refer to as the rebinned cube. Each spatial pixel, 
termed spaxel, in this cube has a size of 9.4~$\times$~9.4~arcsec$^{2}$ 
and thus provides us with a field of view of 47~$\times$~47~arcsec$^{2}$. 
The wavelength regridding parameters \textit{oversample} and 
\textit{upsample} were set to 2 and 1, respectively. This means 
that one spectral bin corresponds to the native resolution of the 
PACS instrument. The integrated [C~II] line flux was obtained by 
spatially integrating the rebinned cube. 

To visualize the extent of the [C~II] emission in Sersic~159-03, we 
created a sky map using the \emph{specProject} task in HIPE and the 
\emph{hrebin} task in IDL\footnote{\url{http://idlastro.gsfc.nasa.gov/ftp/pro/astrom/hrebin.pro}}. We will refer to this cube as the projected 
cube. A pixel size of 6~arcsec was chosen in order to Nyquist 
sample the beam ($\sim$12~arcsec FWHM) at the observed wavelength. 
We only consider spatial bins where the signal-to-noise 
ratio of the integrated line flux is greater than 2. We have compared 
the integrated line fluxes obtained from the rebinned and projected 
cube and find that they are consistent. Velocity and velocity 
width maps were constructed by fitting a single Gaussian to the 
projected cube data.

\begin{table*}[htb]
\centering
\caption{UV and H$\alpha$ Fluxes and Star Formation Rates for S\'ersic~159-03}
\begin{tabular}{l c c c c c c}
\hline\hline
Region & f$_{\textrm{UV}}$ & f$_{\textrm{H}\alpha}$ & SFR$_{\textrm{UV}}^1$ & SFR$_{\textrm{UV}}^2$ & SFR$_{\textrm{H}\alpha}^1$ & SFR$_{\textrm{UV}}$/SFR$_{\textrm{H}\alpha}^1$ \\
 & [10$^{-28}$ erg/s/cm$^{2}$/Hz] & [10$^{-15}$ erg/s/cm$^{2}$] & [M$_{\odot}$ yr$^{-1}$] & [M$_{\odot}$ yr$^{-1}$] & [M$_{\odot}$ yr$^{-1}$] & \\
\hline
Center      &  2.10 $\pm$ 0.04 & 3.82 $\pm$ 0.13 & 0.24 $\pm$ 0.004 & 1.09 $\pm$ 0.02 & 0.24 $\pm$ 0.01 & 0.98\\
North        &  1.17 $\pm$ 0.03 & 9.07 $\pm$ 0.16 & 0.13 $\pm$ 0.003 & 0.60 $\pm$ 0.01 & 0.58 $\pm$ 0.01 & 0.23\\
East          &  0.55 $\pm$ 0.01 & 0.16 $\pm$ 0.08 & 0.06 $\pm$ 0.002 & 0.29 $\pm$ 0.01 & 0.16 $\pm$ 0.01 & 0.40\\
Southeast &  1.65 $\pm$ 0.07 & 0.04 $\pm$ 0.04 & 0.19 $\pm$ 0.007 & 0.85 $\pm$ 0.03 & 0.04 $\pm$ 0.002 & 4.25\\
West         &  0.35 $\pm$ 0.02  & $<$0.50 & 0.04 $\pm$ 0.003 & 0.18 $\pm$ 0.01 & $<$0.03 & $>$9\\
Northwest &  0.03 $\pm$ 0.01 & 0.57 $\pm$ 0.02 & 0.003 $\pm$ 0.001 & 0.015 $\pm$ 0.001 & 0.04 $\pm$ 0.001 & 0.09\\
 & \\
 Total & 5.9 $\pm$ 0.09 & 16.03 $\pm$ 0.23 & 0.66 $\pm$ 0.01 & 3.03 $\pm$ 0.04 & 1.02 & 0.65\\
\hline
\end{tabular}
\tablecomments{All star formation rates assume no additional source of H$\alpha$ ionization or UV emission. All uncertainties are statistical and 1$\sigma$.\\$^1$: Star formation rates from \cite{kennicutt98}, assuming no extinction. \\$^2$: Empirically-calibrated star formation rates from \cite{rg02}, including correction for intrinsic extinction.}
\label{table:sfrs}
\end{table*}

\subsection{Supporting Data}
Throughout the remaining analysis, we compare these new data to previously-published data at a variety of wavelengths. We briefly describe these data below, directing the reader to earlier publications for a more complete description of the data acquisition and reduction.

Narrow-band H$\alpha$ and continuum imaging was presented in \cite{mcdonald10,mcdonald11a} on the central galaxy in S\'ersic~159-03. These data were obtained using the Maryland-Magellan Tunable filter \citep{mmtf} on the Magellan Baade Telescope. The tunable narrow-band filter was centered on rest-frame H$\alpha$ (6563\AA) and 6463\AA, with a width of $\sim$12\AA. These two exposures were differenced in order to make a continuum-free map of H$\alpha$. Due to the narrow width of this filter, some fraction of the total gas may be missing if it is at high/low velocity. These losses are expected to be small, as wider filters do not reveal any additional structures \citep{jaffe05,werner11}. The MMTF H$\alpha$ image revealed a complex, extended ($\sim$35\,kpc) filament to the north of the central cluster galaxy, along with a more narrow filament to the west and an isolated cloud to the northeast. Both the H$\alpha$ and stellar continuum maps are shown in Figure \ref{fig:imtile}.

In addition to the narrow-band continuum image, we use a 300\,s exposure in the Bessel R-band from the Focal Reducer and Spectrograph (FORS1) camera on the Very Large Telescope (VLT). This image has a slightly higher angular resolution than the MMTF continuum image. A more detailed description of these broadband data are provided by \cite{oonk10}.

X-ray data were obtained in 2009 using the Advanced Camera for Surveys (ACIS) on the Chandra X-ray Observatory, and presented in \cite{werner11}. As part of their analysis, \cite{werner11} generated temperature, pressure, and entropy maps, along with unsharp masked images to highlight various substructures, which we utilize here. This earlier work found that the H$\alpha$ emission was preferentially located in regions of high density and low entropy, which we demonstrate in Figure \ref{fig:imtile}. For a complete description of the reduction and analysis of these data, we direct the reader to \cite{million10a,million10b} and \cite{werner11}.

Radio data for the central galaxy in S\'ersic~159-03 is available at a large range of frequencies from 244~MHz to 8.5~GHz. These data are summarized in \cite{werner11}. Here we primarily utilize the 8.5~GHz data from the Very Large Array (VLA) in order to determine the direction of the radio jets. The 8~GHz maps have angular resolution of $3.1^{\prime\prime}\times0.72^{\prime\prime}$, and reveal a powerful radio source centered on the BCG with a jet direction nearly aligned in the north-south direction, as can be seen in Figure \ref{fig:imtile}.

In addition to the data described above, we note the availability of spatially-resolved optical spectroscopy along the two brightest H$\alpha$ filaments from \cite{mcdonald12a} and near-IR spectroscopy from \cite{oonk10}. These data provide kinematics, reddening, and ionization information for the warm ionized and warm molecular gas phases in the extended emission line nebula. While these data are not explicitly utilized in this work, we lean on results from both of these works in our interpretation of the newly-acquired UV and IR emission.

\section{Results}
Below, we summarize new results based on these deep \emph{Herschel} and HST data. We defer a discussion of the greater context of these results to the following section.

\subsection{Young Stellar Populations}

\begin{figure}
\centering
\begin{tabular}{c}
\includegraphics[width=0.42\textwidth]{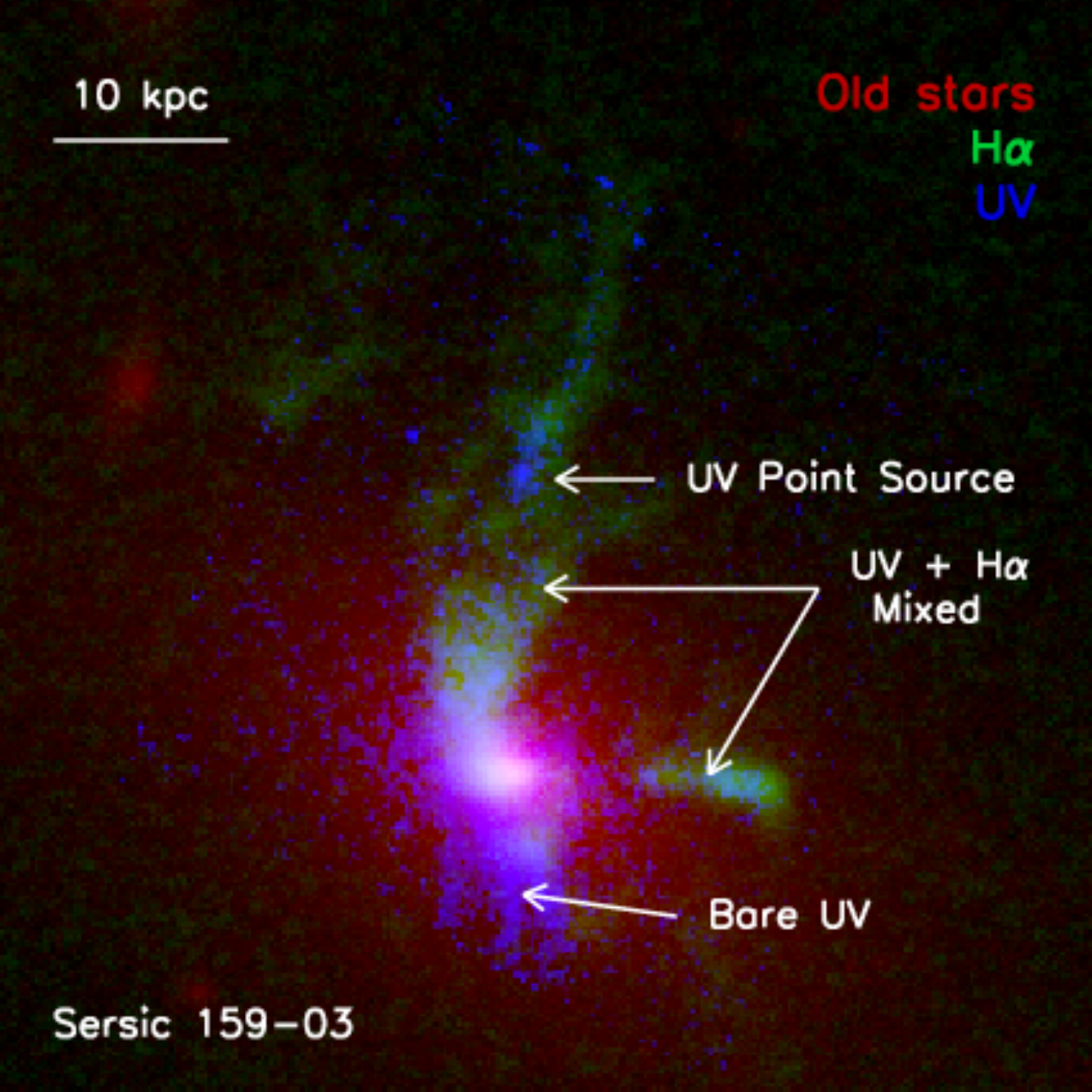}\\
\includegraphics[width=0.42\textwidth]{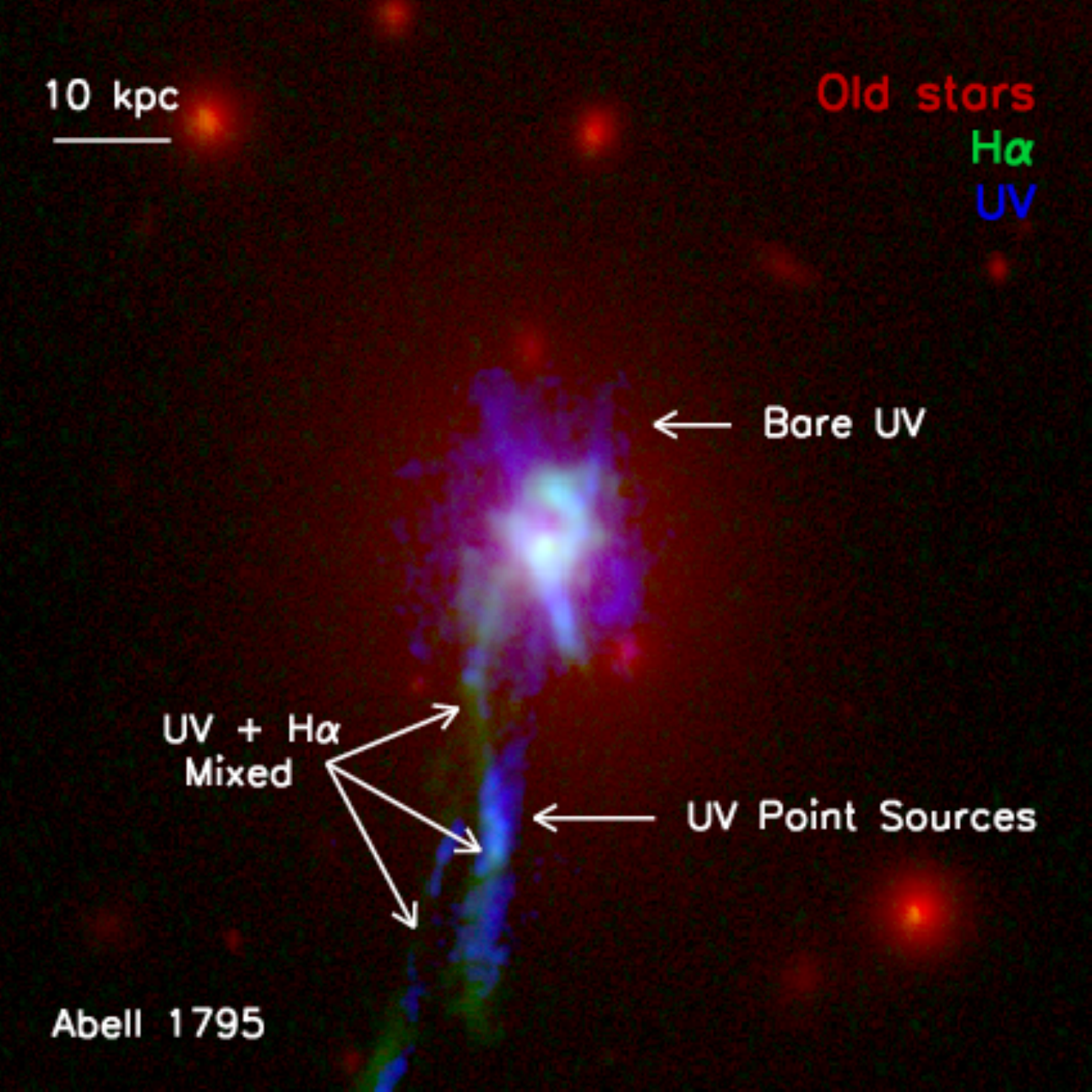}
\end{tabular}
\caption{Three-color images for the central galaxies in S\'ersic~159-03 and Abell~1795 \citep{mcdonald09}. Intriguingly, these two systems exhibit broadly similar traits: in one direction, UV and H$\alpha$ is well mixed along extended filaments, consistent with clumpy, ongoing star formation, while in the opposite direction UV-bright filaments are observed in the absence of any H$\alpha$ emission. }
\label{fig:core}
\end{figure}

We detect extended UV emission both to the north and west of the galaxy center (see Figures \ref{fig:imtile}, \ref{fig:core}), overlapping well with the H$\alpha$ in these regions. The maximal extent of the UV emission is $\sim$30\,kpc to the north and $\sim$15\,kpc to the west. The UV emission to the north and west is clumpy, consistent with what is found in other cool core clusters \citep[e.g.,][]{mcdonald09,odea10}. Roughly 10\% of the total UV emission in the northern filament comes from a single point source -- likely a massive, young star cluster (see discussion section below). As was discussed in \S2.1, the narrow bandpass (F140LP-F150LP) does not contain any bright emission lines, suggesting that this is continuum UV emission. \cite{johnstone12} show that such continuum emission could come from hydrogen two-photon interactions in the particle heating model of \cite{ferland09}. However, this model predicts a ratio of the UV flux to the H$\alpha$ flux of 0.0082, while we measure a value of 0.05 (see Table \ref{table:sfrs}). Correcting for intrinsic extinction would further increase the difference between the model and data in this case, suggesting that particle heating is not producing the UV continuum here. Further, the clumpiness of the UV continuum emission favors a stellar origin. In general, the majority of the H$\alpha$ emission is accompanied by UV emission, with the exception of the faint cloud to the northeast. In addition to being coincident with H$\alpha$ emission, the northern and western UV-bright filaments are coincident with both warm molecular gas \citep{oonk10} and low-entropy intracluster plasma \citep{werner11}. 

To the south and east of the central galaxy, we detect extended UV continuum emission in the absence of H$\alpha$ emission. The UV peak is offset by $\sim$5\,kpc to the south of the central galaxy. This offset may be due to the fact that the dust emission peaks on the central galaxy, which may be obscuring much of the UV emission at this position. Figure \ref{fig:core} highlights the H$\alpha$-free UV emission to the south, with two distinct filaments extending radially in this direction (Figure \ref{fig:core}). These short, H$\alpha$-free filaments are similar to what we observe north of the central galaxy in Abell~1795. Both of these clusters have extended filaments of mixed H$\alpha$ and UV emission in one direction, with ``bare'' UV filaments in the counter direction. This may indicate that recently-formed stars have fallen back onto the central galaxy, with the stars over-shooting the galaxy and the gas accumulating in the center. Given the freefall time at the entropy minimum ($\sim$40 Myr), the youngest stars that make it to the galaxy center ought to be B stars, which may be another explanation for the lack of H$\alpha$ emission. To the east, the UV emission is relatively diffuse and has a similar morphology to the old stellar population (see Figure \ref{fig:imtile}). However, the fact that this emission is not symmetric around the center suggests that this is not UV emission from old stars \citep[e.g.,][]{hicks05}, but rather diffuse emission from a young stellar population.

\begin{figure}
\centering
\includegraphics[width=0.49\textwidth]{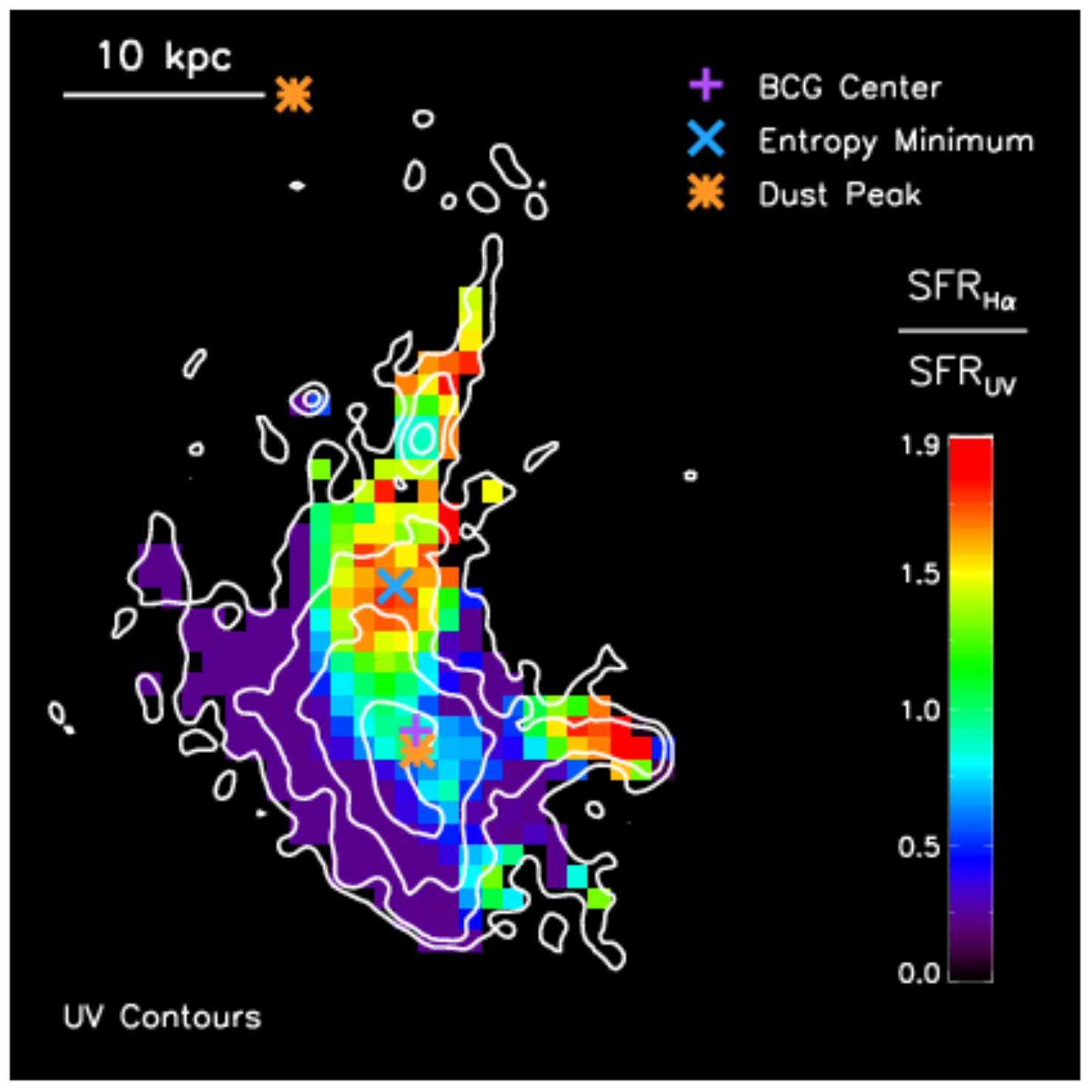}
\caption{H$\alpha$/UV SFR map of emission around the central galaxy in S\'ersic~159-03, where we assume that all H$\alpha$ and UV (F140LP band) emission can be traced to young stars. Assuming that variations in this ratio are due to stellar ages (with young stellar populations having a larger fraction of ionizing O stars), this figure suggests that filaments to the north and west of the central galaxy are sites of ongoing star formation. 
The H$\alpha$/UV distribution peaks to the north of the central galaxy, coincident in position with the lowest entropy X-ray emitting plasma.
The white contours show the extent of the UV emission (see also Figure \ref{fig:imtile}).}
\label{fig:uvha} 
\end{figure}

\begin{figure*}
\centering
\includegraphics[width=0.99\textwidth]{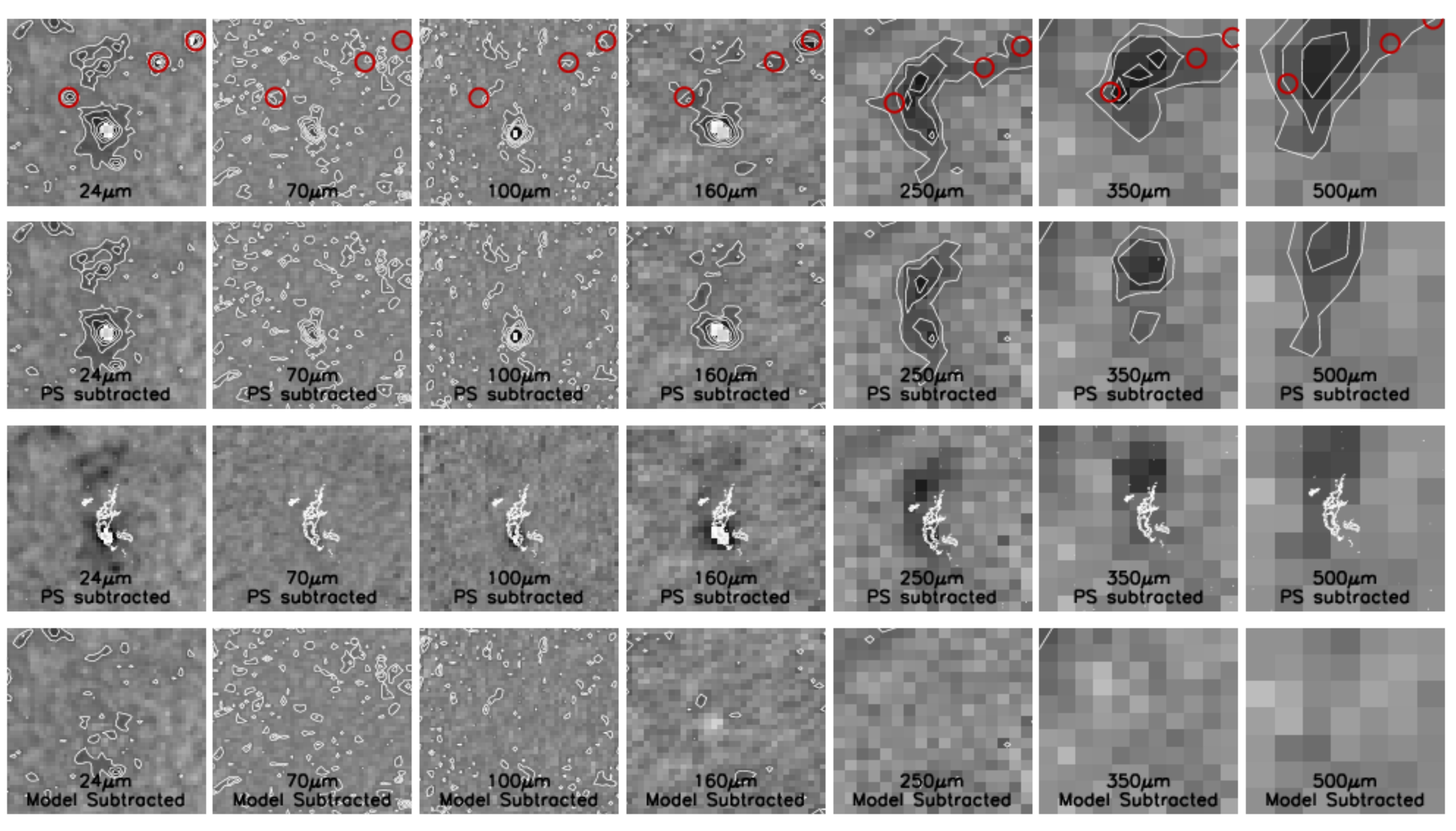}
\caption{IR emission in Spitzer 24$\mu$m and all six PACS+SPIRE bandpasses. All panels are 96$^{\prime\prime}$ (102\,kpc) on a side. In all panels, white contours are at 1$\sigma$, 2$\sigma$, 3$\sigma$, etc. In the upper row we show the raw data, $\sim$2$^{\prime}$ on a side. In the second row (from the top), we show residual images after subtracting point sources unassociated with the central galaxy (marked with red circles). At all wavelengths, we see evidence of residual, extended emission. This emission can not be removed with the addition of a single point source, suggesting that it is indeed extended. In the third row, we show the same residual image, but with H$\alpha$ contours overlaid. The brightness of the residual IR emission at long wavelengths suggests that it may be a cloud/filament of cold dust (T$_{dust} \sim 12$K; see \S3.3, Fig.\ \ref{fig:irsed}), coincident in position with the most extended H$\alpha$ emission. In the bottom panel we show the residual after our model for the extended dust and central region has been subtracted, which demonstrates that we are capturing all of the extended emission.}
\label{fig:ir}
\end{figure*}

In Table \ref{table:sfrs} we show the integrated UV flux in the F140LP passband for various morphologically-defined regions. The total UV flux in this system is 5.9 $\pm$ 0.09 $\times$10$^{-28}$ erg s$^{-1}$ cm$^{-2}$ Hz$^{-1}$. Assuming no intrinsic extinction and that all of the UV flux comes from young stars, this corresponds to a star formation rate (SFR) of 0.66 M$_{\odot}$ yr$^{-1}$, following \cite{kennicutt98}. This is most likely an underestimate of the total SFR, given that we measure significant reddening via optical spectroscopy \citep{mcdonald12a} and IR continuum emission (Figure \ref{fig:imtile}). Assuming an empirically-calibrated relation between intrinsic extinction and UV luminosity \citep{rg02}, we estimate an extinction-corrected star formation rate of $\sim3$ M$_{\odot}$ yr$^{-1}$. This is slightly higher than the H$\alpha$-derived star formation rate of $\sim$1 M$_{\odot}$ yr$^{-1}$, where we have again assumed that the H$\alpha$ flux is entirely due to photo-ionization by young stars. This difference may be, in part, due to the narrow bandpass of the MMTF -- \cite{werner11} quote an H$\alpha$ flux 35\% higher than our measured value, using a wider bandpass. Unfortunately, this wider bandpass is contaminated by emission from the  [N\,\textsc{ii}] doublet, meaning that the true H$\alpha$ flux most likely lies somewhere between these two measurements.

We measure a broad range of UV/H$\alpha$ ratios across the full extent of the central galaxy, suggesting that simple continuous star formation is insufficient to explain the UV and H$\alpha$ morphologies. To the west and southeast of the galaxy center, there is a dearth of H$\alpha$ emission. This can not be explained by extinction or by including an additional source of ionization. The likely explanation is that the stellar populations in these regions are older than in the more extended filaments. If we carry this explanation further and assume that the H$\alpha$/UV ratio probes the age of the young stars -- where recently-formed stars have an abundance of ionizing O stars, and older stellar populations have their UV flux dominated by weakly-ionizing B stars -- we can make a pseudo-age map by simply dividing these two images. This is demonstrated in Figure \ref{fig:uvha}. We find that the H$\alpha$/UV ratio peaks to the north of the central galaxy, precisely where the ICM entropy is minimized. Given that the IR emission is relatively faint at this position (see Figure \ref{fig:imtile}), we expect that this high ratio is not a result of intrinsic extinction. Instead, we propose that this northern region, along with the bright western filament, are sites of ongoing star formation. In contrast, the southern and western regions have low H$\alpha$/UV ratios, suggesting an older stellar population. We will return to this discussion of stellar ages in \S4.

\subsection{IR Morphology -- Extended Gas and Dust Emission}
In Figure \ref{fig:imtile} we show that the 250$\mu$m emission peaks on the central cluster galaxy and extends to the northwest, out to the maximum extent of the H$\alpha$ filaments. In Figure \ref{fig:ir} we show that this dust emission is present at all wavelengths $\ge$24$\mu$m, while the central galaxy is brightest at $<$250$\mu$m. Modeling and subtracting emission from three nearby point sources, we can isolate the extended dust emission. This emission is peaked just north of the H$\alpha$ filament. The brightness of this dust at such long wavelengths suggests that it is cold ($\sim$12\,K; see \S3.3), if it is located at the distance of S\'ersic~159-03. The amount of dust in the extended region to the north and northeast is sufficient to prevent the detection of any UV emission from this region, perhaps explaining why the UV emission is not as extended in this direction as the H$\alpha$ emission.

\subsection{Dust Temperature \& Mass}
\begin{figure}
\centering
\vspace{1.3cm}
\begin{minipage}{0.49\textwidth}
\includegraphics[width=8.2cm,trim=0.5cm 9.5cm 2cm 6.5cm]{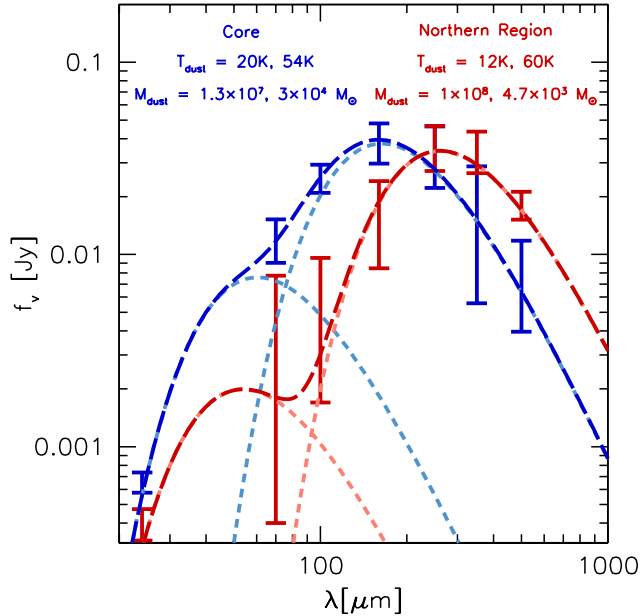}
\end{minipage}
\vspace{2cm}
\caption{IR spectral energy distribution (SED) for the central (blue) and extended (red) emission shown in Figure \ref{fig:ir}. We fit both distributions with two-temperature modified blackbody models, assuming $\beta=1.5$. We find that the bulk of the dust is cold, with a modest warm (54\,K) component, which is likely heated by the central AGN. The total mass of dust in this system is $\sim$10$^8$ M$_{\odot}$, on par with other cool core clusters.}
\label{fig:irsed}
\end{figure}

The relative lack of extended emission at shorter wavelengths suggests that the dust to the north of the galaxy is significantly cooler than that in the center. This is confirmed in Figure \ref{fig:irsed}, which shows the IR spectral energy distributions (SEDs) for the core emission (centered on the BCG) and the extended emission to the north of the central galaxy. A pair of modified blackbody models were fit to these SEDs, assuming $\beta=1.5$ and normalizing the absorption cross section to $\kappa_{abs}$(250$\mu$m) = 4.0 cm$^2$ g$^{-1}$ \citep{weingartner01}. We note that this normalization term is known to be quite uncertain \citep[e.g.,][]{bianchi13}. In the inner $\sim$20\,kpc, the dust has a mix of temperatures with a total dust mass of $\sim$10$^7$ M$_{\odot}$. The majority of the flux comes from a cool 20\,K component, likely associated with the star formation in the central galaxy. A second, warmer (54\,K), component in the core is likely attributed to the central AGN. These two dust temperatures are consistent with typical values for star-forming galaxies and central AGN \citep[e.g.,][]{clemens13}.

The extended emission originates predominantly from cold dust (12\,K), with M$_{dust} \sim10^8$\,M$_{\odot}$. Interestingly, this suggests that the vast majority of the dust, by mass, is at large radii ($\gtrsim$30\,kpc). The total dust mass in this system is similar to that observed in Abell\,1068 and Zw\,3146 \citep{edge10b}, despite the fact that these clusters have star formation rates more than a factor of ten times higher than that measured in the center of S\'ersic~159-03. We measure a small secondary peak at $\sim$60\,K, suggesting that some of the dust in this cloud is being heated, perhaps by interaction with the ICM.

Considering only the cool dust in the central region, where we see evidence for star formation, the total infrared luminosity is L$_{\textrm{IR}} = 1.8\times10^{10}$ L$_{\odot}$, corresponding to a star formation rate of 2.3 M$_{\odot}$ yr$^{-1}$, following \citep{bell03}. The total infrared luminosity from cool dust in the extended region is L$_{\textrm{IR}} = 1.0\times10^{10}$ L$_{\odot}$, suggesting that there may be an additional $\sim$1 M$_{\odot}$ yr$^{-1}$ in obscured star formation in the extended northern filament.

\subsection{Cold gas}
Figure \ref{fig:imtile} demonstrates that [C\,\textsc{ii}] 157$\mu$m emission is present along the full extent of the UV/H$\alpha$ filaments, suggesting that there is a substantial amount of cold ($\sim$100\,K) gas in the core of S\'ersic~159-03. Similar to the UV emission, the cool gas is significantly more extended to the south and west of the central galaxy than the H$\alpha$ emission. We measure an integrated [C\,\textsc{ii}] flux of $2.3 \pm 0.5 \times 10^{-14}$ erg s$^{-1}$ cm$^{-2}$, corresponding to a luminosity of 
$1.9 \pm 0.4 \times 10^{41}$ erg s$^{-1}$. Roughly 40\% of this emission is coming from the central $9.4^{\prime\prime}\times9.4^{\prime\prime}$ spaxel. Following \cite{herrera-camus14}, we estimate a SFR derived from the [C\,\textsc{ii}] of $0.95 \pm 0.2$ M$_{\odot}$ yr$^{-1}$, which is within a factor of 2 of the H$\alpha$-, UV-, and FIR-derived estimates.

\begin{figure}
\centering
\includegraphics[width=0.45\textwidth]{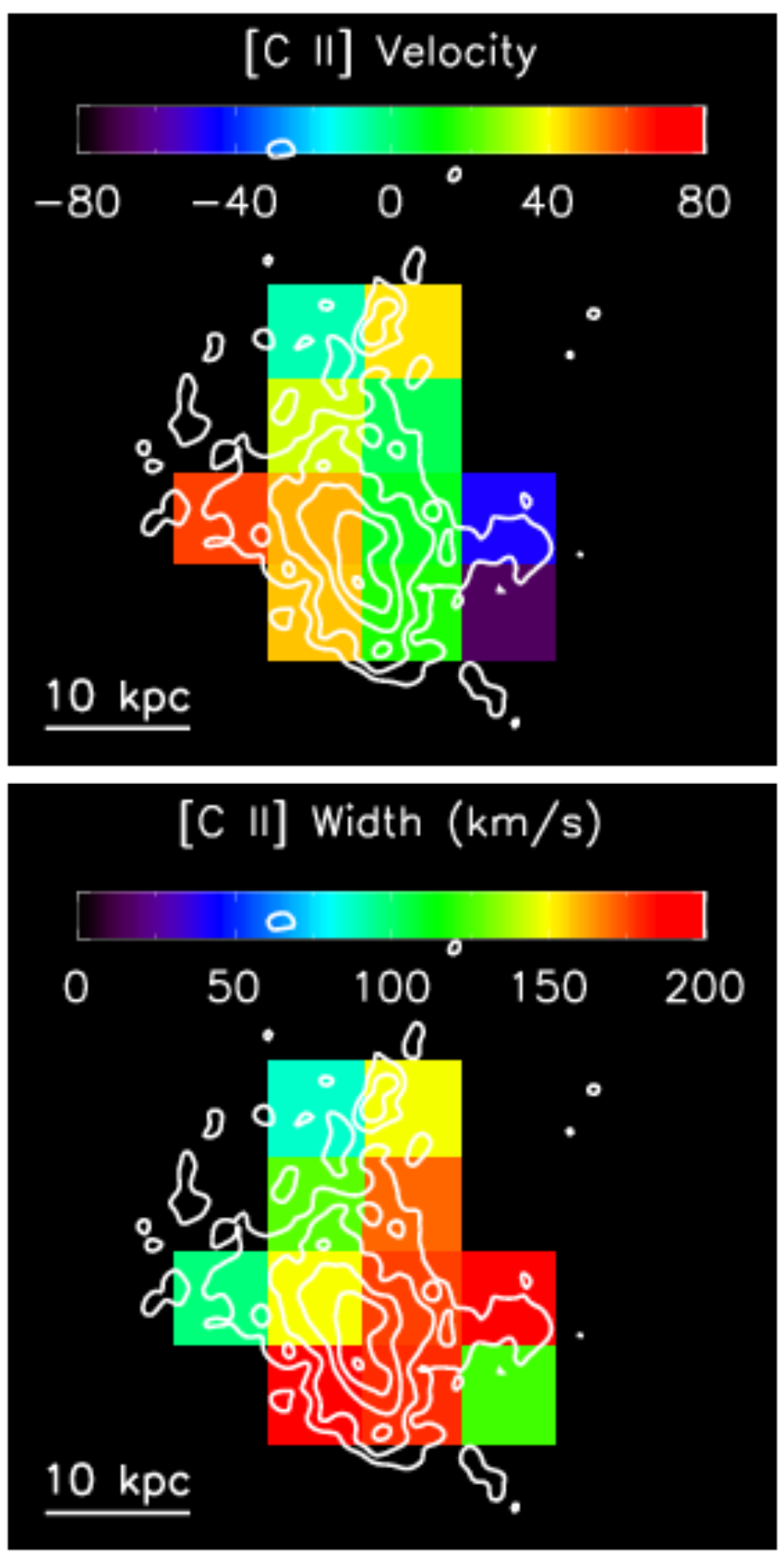}
\caption{Kinematic information from PACS spectroscopy. In the upper panel we show the line-of-sight velocity of the [C\,\textsc{ii}]-emitting gas, while the lower panel shows the velocity width of the emission line. The kinematics of the cool gas are broadly consistent with the warm, molecular \citep{oonk10} and ionized gas \citep{mcdonald12a}.}
\label{fig:cii}
\end{figure}

The pixel-to-pixel detection of [C\,\textsc{ii}] emission is only marginally significant, preventing us from performing a detailed kinematic analysis of the cold gas. Broadly speaking (Figure \ref{fig:cii}) the kinematics of the cold gas appear consistent with those of the warm molecular \citep{oonk10} and ionized \citep{mcdonald12a} gas. The velocity dispersion peaks at a value of $\sim$200 km s$^{-1}$ toward the galaxy center, dropping to $\sim$100 km s$^{-1}$ in the more extended filaments. There is very little variation in the line-of-sight velocity along the extended filaments ($\Delta v$ $\sim$ 50 km s$^{-1}$), suggesting that the filaments lie (primarily) in the plane of the sky.

\section{Discussion}
Below, we discuss possible interpretations of the UV, IR, and [C\,\textsc{ii}]\,157$\mu$m emission presented in \S3. This is, by no means, an exhaustive list of possible contributing sources, but is meant to provide a broad, coherent picture of what may be occurring in the core of S\'ersic~159-03.

\subsection{Star Formation in Condensing, Low-Entropy ICM}

In \S3 and previous works \citep{mcdonald10,mcdonald12a,werner11,oonk10} we show significant evidence for star formation in the central galaxy of S\'ersic~159-03. The measured star formation rate of $\sim$1--3 M$_{\odot}$ yr$^{-1}$ (derived from UV, H$\alpha$, far-IR, and [C\,\textsc{ii}]) is concentrated in the center of the galaxy, but extends in all directions for $\sim$10--15\,kpc and to the north for $\sim$35\,kpc. \cite{werner11} showed that the extended northern filament is coincident with an offset cool core, suggesting that star formation may be triggered by whatever process has dislodged the cool core from the minimum of the gravitational potential.

%4pid^2 = 7.18e54
% P = 0.012 keV cm^-3
% V = 3.32e69 cm^3 (30 kpc)
% 4pV = 2.6e59
Sloshing of the cool core following an interaction with another massive system is the most common explanation for offset cool cores \citep[e.g.,][]{zuhone10,zuhone11,blanton11,ehlert14}.
The large-scale X-ray emission of the ICM is, however, extremely smooth in this system \citep{werner11}, with no obvious indications of sloshing (i.e., cold fronts, spiral structure, asymmetric isophotes). An alternative explanation, for which there is some evidence, is that a burst of radio-mode feedback has dislodged the cool core to the north. This hypothesis is consistent with the direction of the radio jet (Figure \ref{fig:imtile}) which is aligned along the north-south direction. Following the strategy described in \cite{sun09b}, we can estimate whether or not the current AGN power is sufficient to displace the cool core. Requiring an energy outburst of $4\int PdV$ to remove the cool core yields an estimate of $8\times10^{43}$ erg s$^{-1}$, assuming a timescale of 10$^8$ yr ($3.16\times10^{15}$ s), a cool core radius of 30\,kpc ($V = 3.3\times10^{69}$ cm$^3$), and a core pressure of 0.012 keV cm$^{-3}$ \citep[$1.92\times10^{-11}$ erg cm$^{-3}$;][]{werner11}. Following \cite{birzan08}, and using the measured 1.4\,GHz luminosity of $1.7\times10^{31}$ erg s$^{-1}$ Hz$^{-1}$ \citep{werner11}, we estimate a mechanical power of $\sim$10$^{44}$erg s$^{-1}$ from the radio-loud AGN. Thus, it appears that the current outburst of AGN feedback is sufficient, both in power and direction, to displace the cool core to its current, observed position.

In this scenario, the hot gas is displaced most easily by the radio-mode feedback -- the molecular gas, warm gas, and dust are not as easily pushed around by the AGN jets. We speculate then that the filaments of young stars and ionized gas are, instead, due to recent star formation as the cool core moves from the center to the north -- not unlike a passing raincloud leaving everything in its path wet. \citet{werner11} estimate that the age of the extended filament is $\sim$10$^8$~yr. Removed from the direct influence of the central AGN, and compressed by the AGN jets/lobes, the displaced cool core should cool rapidly. This scenario is corroborated by the peak in the H$\alpha$/UV ratio at \emph{precisely} the location of the entropy minimum. This seems to indicate that the youngest, strongly-ionizing O stars are located at the entropy minimum, \emph{regardless of whether or not that minimum lies in the cluster center}. There are, of course, other possible explanations for a high H$\alpha$/UV ratio. 
Reddening due to dust would reduce the observed UV flux considerably more so than the H$\alpha$ flux.
However, both the dust map (Figure \ref{fig:ir}) and the reddening map \citep{mcdonald12a} show minimal dust at the position of the entropy minimum. Alternatively, the H$\alpha$/UV ratio could be increased by including a secondary ionization source such as shocks \citep[e.g.,][]{mcdonald12a}, particle heating \citep[e.g.,][]{ferland09}, or mixing with the hot ICM \citep[e.g.][]{fabian11}. We note that both the [O\,\textsc{i}]/H$\alpha$ ratio and the velocity dispersion are minimized at the position of the entropy minimum, suggesting that turbulent mixing and/or shocks are probably not strong contributors. Given that the spectral signatures of cooling and mixing are, for the most part, indistinguishable, it is difficult to rule out the possibility that the H$\alpha$ may peak due to the cooling enhancement at this position. In either case (cooling, mixing, star formation), the H$\alpha$/UV peak appears to be signaling the presence of rapid cooling at the position of the entropy minimum, leading to the formation of the stars in the northern filament.

Following these same lines of arguments, the low values of the H$\alpha$/UV ratio in the center, south, and western regions would indicate that the stars here are slightly older. These stars were likely formed when the cool core was still centered on the BCG, and the cooling/feedback loop was closed. The recent burst of AGN feedback has disrupted this loop, halting star formation in the central regions and relocating it to the northern filament. 

Star formation from extended filaments produced by AGN uplift may contribute significantly to the formation of the outer stellar haloes and massive star-clusters surrounding cD galaxies \citep{canning10,canning14}. About 10\% of the observed UV emission in the northern filament of S\'ersic~159-03 comes from a single massive, young star cluster (see Figure \ref{fig:core}). This is consistent with the star forming filaments in the outskirts of the Perseus Cluster, where 10\% of the young star light is in star-clusters \citep{canning14}. 

Shock heating and compressing the cold gas by a galaxy falling through the core of the cluster might also have triggered and enhanced the star formation rate in the filament \citep[e.g.][]{roediger14}. Optical images of the cluster core show a pair of small, perturbed galaxies $\sim$30--40~kpc northeast of the cD galaxy (see Figure \ref{fig:fors1}). It is unclear whether either of these galaxies are cluster members (we lack redshift information), although their apparent size and magnitude are appropriate given the distance to Sersic~159-03. If these galaxies passed through the cluster core (the core passage would have occurred $\sim$60~Myr ago) they might have interacted with the cold gas in the cluster center, triggering some of the observed star formation and contributing to the disturbed distribution of the cold gas. Further, some of the warm, ionized gas, specifically in the northeast filament, may have been stripped from the satellite galaxy during this interaction. Given the morphology of the bulk of the H$\alpha$ and UV emission (corresponding well with the X-ray morphology), we do not expect that a significant fraction of the gas or stars were stripped from these galaxies, but their passage may aid in the triggering of star formation and the ``stretching out'' of additional filaments via tidal interactions.

\subsection{Origin of Dusty Filament}

We find a significant amount of extended, cold dust to the north of the central galaxy in S\'ersic~159-03 (Figure \ref{fig:ir}). This massive (10$^8$ M$_{\odot}$ yr$^{-1}$) cloud of cold (12\,K) dust extends to the north of the central galaxy for $\gtrsim$30\,kpc, making it the most extended complex of dust yet found in a cool core cluster. If truly associated with the star-forming filaments, it suggests that these filaments extend further to the north than is observed, and would explain why the UV emission is less extended than the H$\alpha$ emission. There is, however, the possibility that this dust is associated with background galaxies, in an unfortunate alignment with both the central galaxy and the extended filament.

\begin{figure}
\centering
\includegraphics[width=0.45\textwidth]{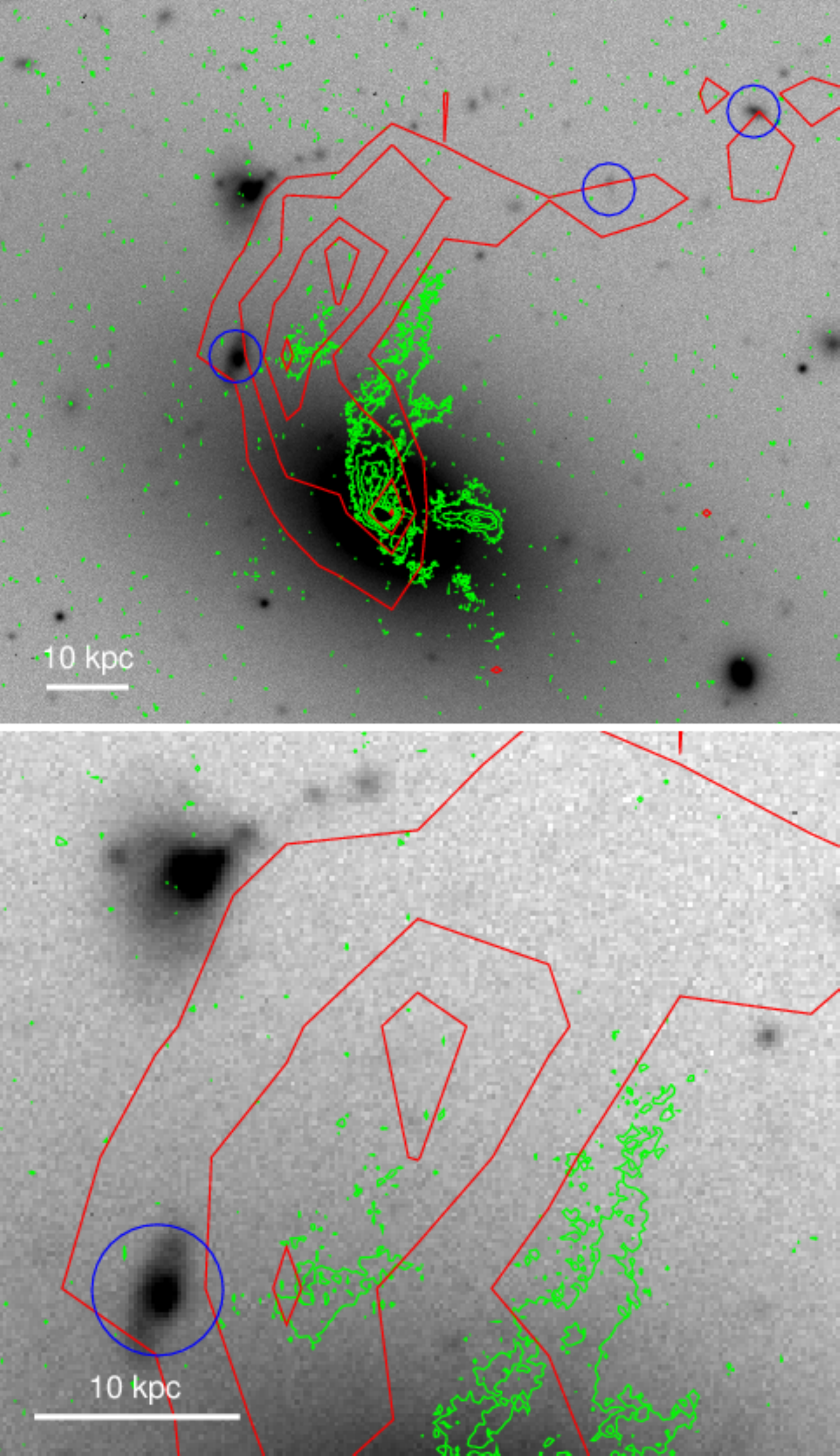}
\caption{Deep R-band image of the central cluster galaxy in S\'ersic~159-03, from VLT-FORS1 (see \S2.4). In both panels, red contours show 250$\mu$m emission, green contours show H$\alpha$ emission, and blue circles show 24$\mu$m sources which we have subtracted in our analysis. In the lower panel we zoom in and center on the cold dust peak to the north of the central galaxy, showing a lack of background galaxies that could be contributing to this flux.}
\label{fig:fors1}
\end{figure}

Attributing this dust to background galaxies would help alleviate the issue of sustaining such a large quantity of cold dust surrounded by hot intracluster gas. Deep optical imaging along the 250\,$\mu$m filament reveals three galaxies which may be contributing to the IR flux (Figure \ref{fig:fors1}) and are coincident with peaks in the extended northern filament. Emission from these sources were subtracted and removed (see Figure \ref{fig:ir}). In the residual IR peak, highlighted in Figure \ref{fig:fors1}, we see no evidence for background galaxies. This would, conservatively, put a lower limit of $z\gtrsim1$ on 0.1L$_*$ galaxies that may be contributing to the cold dust peak to the north of the central galaxy in Sersic~159-03. If this IR emission did, indeed, come from galaxies at $z>1$ it would imply L$_{\textrm{IR}} > 7.5\times10^{12}$ L$_{\odot}$, which is a factor of $\sim$10 times larger than the typical luminosity of an ``ultraluminous infrared galaxy'' \citep{gao99,klaas01}.

In the absence of an alternative, viable explanation for the source of the dust, we conclude that it is most likely associated with the northern star-forming filament. It is unclear how dust this cold (12\,K) can avoid being heated by the hot ICM, but the fact that we see a warm (60\,K) component in the extended emission suggests that the dust cloud may be in the process of heating up along the outer surface. Future observations with ALMA will allow us to both confirm that the dust is indeed linked to the cooling filament, and look for cold molecular gas along its full extent.

\section{Summary}
Cooling and star formation in the cores of clusters of galaxies appear to be regulated by feedback from the central AGN.
Here, we studied the properties of the dusty, filamentary cold gas and star formation in the core of the galaxy cluster S\'ersic 159-03 using new FUV images from HST and FIR data from \emph{Herschel}, along with a wealth of archival data.
We find:

\begin{itemize}

\item Extended, filamentary and clumpy UV emission, indicating a star formation rate of $\sim1$--3~$M_{\odot}$~yr$^{-1}$ and an abundance ($\sim10^7~M_\odot$) of cold (20~K) dust in the center of the cluster. The morphology of the UV, H$\alpha$, [C\,\textsc{ii}], and IR emission are all broadly consistent with one another.

\item  The most recently formed stars, indicated by the peak H$\alpha$/UV ratio, are spatially coincident with the lowest entropy ICM, that has been displaced by the AGN $\sim$7.5\,kpc north of the center of the cD galaxy. This suggests that the displacement of the cooling core from the direct vicinity of the central AGN can temporarily break the feedback cycle and lead to cooling and star formation that is offset from the center of the galaxy. 

\item 
Extended FIR continuum emission is observed north to north-east of the central galaxy. If this FIR emission is located at the distance of the cluster then this implies that large amounts ($\sim$10$^8$\,M$_\odot$) of cold (12\,K) dust may be associated with the H$\alpha$ filament $\sim$30\,kpc north of the central galaxy. This would be the most extended complex of dust yet found in a cool core cluster. If associated with the cooling flow, this would imply an additional $\sim$1 M$_{\odot}$ yr$^{-1}$ of obscured star formation.
\end{itemize}

These results, and our interpretation, could be corroborated by high angular resolution sub-mm observations, both of cold molecular gas and far-IR continuum. If the ``focus'' of cooling has really been relocated from the cluster center, then we would expect the cold gas reservoir to be depleted on the central galaxy, and enhanced to the north. 

\section*{Acknowledgements} 
{\it Herschel} is an ESA space observatory with science instruments provided by European-led Principal Investigator consortia and with important participation from NASA.
M. M. acknowledges support by NASA through a Hubble Fellowship grant HST-HF51308.01-A awarded by the Space Telescope Science Institute, which is operated by the Association of Universities for Research in Astronomy, Inc., for NASA, under contract NAS 5-26555. We also acknowledge support from grant HST-GO-12570.01-A. NW thanks Rebecca Canning for helpful discussions.
%

%\bibliographystyle{apj}
%\bibliography{ref}

\begin{thebibliography}{}
\expandafter\ifx\csname natexlab\endcsname\relax\def\natexlab#1{#1}\fi

\bibitem[{{Balog} {et~al.}(2014){Balog}, {M{\"u}ller}, {Nielbock}, {Altieri},
  {Klaas}, {Blommaert}, {Linz}, {Lutz}, {Mo{\'o}r}, {Billot}, {Sauvage}, \&
  {Okumura}}]{balog14}
{Balog}, Z., {M{\"u}ller}, T., {Nielbock}, M., {et~al.} 2014, Experimental
  Astronomy, 37, 129

\bibitem[{{Bell}(2003)}]{bell03}
{Bell}, E.~F. 2003, \apj, 586, 794

\bibitem[{{Bendo} {et~al.}(2013){Bendo}, {Griffin}, {Bock}, {Conversi},
  {Dowell}, {Lim}, {Lu}, {North}, {Papageorgiou}, {Pearson}, {Pohlen},
  {Polehampton}, {Schulz}, {Shupe}, {Sibthorpe}, {Spencer}, {Swinyard},
  {Valtchanov}, \& {Xu}}]{bendo13}
{Bendo}, G.~J., {Griffin}, M.~J., {Bock}, J.~J., {et~al.} 2013, \mnras, 433,
  3062

\bibitem[{{Bianchi}(2013)}]{bianchi13}
{Bianchi}, S. 2013, \aap, 552, A89

\bibitem[{{B{\^i}rzan} {et~al.}(2008){B{\^i}rzan}, {McNamara}, {Nulsen},
  {Carilli}, \& {Wise}}]{birzan08}
{B{\^i}rzan}, L., {McNamara}, B.~R., {Nulsen}, P.~E.~J., {Carilli}, C.~L., \&
  {Wise}, M.~W. 2008, \apj, 686, 859

\bibitem[{{Blanton} {et~al.}(2011){Blanton}, {Randall}, {Clarke}, {Sarazin},
  {McNamara}, {Douglass}, \& {McDonald}}]{blanton11}
{Blanton}, E.~L., {Randall}, S.~W., {Clarke}, T.~E., {et~al.} 2011, \apj, 737,
  99

\bibitem[{{Canning} {et~al.}(2010){Canning}, {Fabian}, {Johnstone}, {Sanders},
  {Conselice}, {Crawford}, {Gallagher}, \& {Zweibel}}]{canning10}
{Canning}, R.~E.~A., {Fabian}, A.~C., {Johnstone}, R.~M., {et~al.} 2010,
  \mnras, 405, 115

\bibitem[{{Canning} {et~al.}(2014){Canning}, {Ryon}, {Gallagher}, {Kotulla},
  {O'Connell}, {Fabian}, {Johnstone}, {Conselice}, {Hicks}, {Rosario}, \&
  {Wyse}}]{canning14}
{Canning}, R.~E.~A., {Ryon}, J.~E., {Gallagher}, J.~S., {et~al.} 2014, \mnras,
  444, 336

\bibitem[{{Clemens} {et~al.}(2013){Clemens}, {Negrello}, {De Zotti},
  {Gonzalez-Nuevo}, {Bonavera}, {Cosco}, {Guarese}, {Boaretto}, {Salucci},
  {Baccigalupi}, {Clements}, {Danese}, {Lapi}, {Mandolesi}, {Partridge},
  {Perrotta}, {Serjeant}, {Scott}, \& {Toffolatti}}]{clemens13}
{Clemens}, M.~S., {Negrello}, M., {De Zotti}, G., {et~al.} 2013, \mnras, 433,
  695

\bibitem[{{Crawford} {et~al.}(1999){Crawford}, {Allen}, {Ebeling}, {Edge}, \&
  {Fabian}}]{crawford99}
{Crawford}, C.~S., {Allen}, S.~W., {Ebeling}, H., {Edge}, A.~C., \& {Fabian},
  A.~C. 1999, \mnras, 306, 857

\bibitem[{{de Plaa} {et~al.}(2006){de Plaa}, {Werner}, {Bykov}, {Kaastra},
  {M{\'e}ndez}, {Vink}, {Bleeker}, {Bonamente}, \& {Peterson}}]{deplaa2006}
{de Plaa}, J., {Werner}, N., {Bykov}, A.~M., {et~al.} 2006, \aap, 452, 397

\bibitem[{{Donahue} {et~al.}(2000){Donahue}, {Mack}, {Voit}, {Sparks},
  {Elston}, \& {Maloney}}]{donahue00}
{Donahue}, M., {Mack}, J., {Voit}, G.~M., {et~al.} 2000, \apj, 545, 670

\bibitem[{{Edge}(2001)}]{edge01}
{Edge}, A.~C. 2001, \mnras, 328, 762

\bibitem[{{Edge} \& {Frayer}(2003)}]{edge03}
{Edge}, A.~C., \& {Frayer}, D.~T. 2003, \apjl, 594, L13

\bibitem[{{Edge} {et~al.}(2002){Edge}, {Wilman}, {Johnstone}, {Crawford},
  {Fabian}, \& {Allen}}]{edge02}
{Edge}, A.~C., {Wilman}, R.~J., {Johnstone}, R.~M., {et~al.} 2002, \mnras, 337,
  49

\bibitem[{{Edge} {et~al.}(2010{\natexlab{a}}){Edge}, {Oonk}, {Mittal}, {Allen},
  {Baum}, {B{\"o}hringer}, {Bregman}, {Bremer}, {Combes}, {Crawford},
  {Donahue}, {Egami}, {Fabian}, {Ferland}, {Hamer}, {Hatch}, {Jaffe},
  {Johnstone}, {McNamara}, {O'Dea}, {Popesso}, {Quillen}, {Salom{\'e}},
  {Sarazin}, {Voit}, {Wilman}, \& {Wise}}]{edge10a}
{Edge}, A.~C., {Oonk}, J.~B.~R., {Mittal}, R., {et~al.} 2010{\natexlab{a}},
  \aap, 518, L46

\bibitem[{{Edge} {et~al.}(2010{\natexlab{b}}){Edge}, {Oonk}, {Mittal}, {Allen},
  {Baum}, {B{\"o}hringer}, {Bregman}, {Bremer}, {Combes}, {Crawford},
  {Donahue}, {Egami}, {Fabian}, {Ferland}, {Hamer}, {Hatch}, {Jaffe},
  {Johnstone}, {McNamara}, {O'Dea}, {Popesso}, {Quillen}, {Salom{\'e}},
  {Sarazin}, {Voit}, {Wilman}, \& {Wise}}]{edge10b}
---. 2010{\natexlab{b}}, \aap, 518, L47

\bibitem[{{Edwards} {et~al.}(2007){Edwards}, {Hudson}, {Balogh}, \&
  {Smith}}]{edwards07}
{Edwards}, L.~O.~V., {Hudson}, M.~J., {Balogh}, M.~L., \& {Smith}, R.~J. 2007,
  \mnras, 379, 100

\bibitem[{{Ehlert} {et~al.}(2014){Ehlert}, {McDonald}, {Miller}, {David}, \&
  {Bautz}}]{ehlert14}
{Ehlert}, S., {McDonald}, M., {Miller}, E.~D., {David}, L.~P., \& {Bautz},
  M.~W. 2014, ArXiv e-prints, arXiv:1406.4352

\bibitem[{{Ehlert} {et~al.}(2011){Ehlert}, {Allen}, {von der Linden},
  {Simionescu}, {Werner}, {Taylor}, {Gentile}, {Ebeling}, {Allen}, {Applegate},
  {Dunn}, {Fabian}, {Kelly}, {Million}, {Morris}, {Sanders}, \&
  {Schmidt}}]{ehlert11}
{Ehlert}, S., {Allen}, S.~W., {von der Linden}, A., {et~al.} 2011, \mnras, 411,
  1641

\bibitem[{{Fabian} {et~al.}(2011){Fabian}, {Sanders}, {Williams}, {Lazarian},
  {Ferland}, \& {Johnstone}}]{fabian11}
{Fabian}, A.~C., {Sanders}, J.~S., {Williams}, R.~J.~R., {et~al.} 2011, \mnras,
  417, 172

\bibitem[{{Falcke} {et~al.}(1998){Falcke}, {Rieke}, {Rieke}, {Simpson}, \&
  {Wilson}}]{falcke98}
{Falcke}, H., {Rieke}, M.~J., {Rieke}, G.~H., {Simpson}, C., \& {Wilson}, A.~S.
  1998, \apjl, 494, L155

\bibitem[{{Farage} {et~al.}(2012){Farage}, {McGregor}, \& {Dopita}}]{farage12}
{Farage}, C.~L., {McGregor}, P.~J., \& {Dopita}, M.~A. 2012, \apj, 747, 28

\bibitem[{{Ferland} {et~al.}(2009){Ferland}, {Fabian}, {Hatch}, {Johnstone},
  {Porter}, {van Hoof}, \& {Williams}}]{ferland09}
{Ferland}, G.~J., {Fabian}, A.~C., {Hatch}, N.~A., {et~al.} 2009, \mnras, 392,
  1475

\bibitem[{{Gao} \& {Solomon}(1999)}]{gao99}
{Gao}, Y., \& {Solomon}, P.~M. 1999, \apjl, 512, L99

\bibitem[{{Gaspari} {et~al.}(2012){Gaspari}, {Ruszkowski}, \&
  {Sharma}}]{gaspari12}
{Gaspari}, M., {Ruszkowski}, M., \& {Sharma}, P. 2012, \apj, 746, 94

\bibitem[{{Griffin} {et~al.}(2010){Griffin}, {Abergel}, {Abreu}, {Ade},
  {Andr{\'e}}, {Augueres}, {Babbedge}, {Bae}, {Baillie}, {Baluteau}, {Barlow},
  {Bendo}, {Benielli}, {Bock}, {Bonhomme}, {Brisbin}, {Brockley-Blatt},
  {Caldwell}, {Cara}, {Castro-Rodriguez}, {Cerulli}, {Chanial}, {Chen},
  {Clark}, {Clements}, {Clerc}, {Coker}, {Communal}, {Conversi}, {Cox},
  {Crumb}, {Cunningham}, {Daly}, {Davis}, {de Antoni}, {Delderfield}, {Devin},
  {di Giorgio}, {Didschuns}, {Dohlen}, {Donati}, {Dowell}, {Dowell}, {Duband},
  {Dumaye}, {Emery}, {Ferlet}, {Ferrand}, {Fontignie}, {Fox}, {Franceschini},
  {Frerking}, {Fulton}, {Garcia}, {Gastaud}, {Gear}, {Glenn}, {Goizel},
  {Griffin}, {Grundy}, {Guest}, {Guillemet}, {Hargrave}, {Harwit}, {Hastings},
  {Hatziminaoglou}, {Herman}, {Hinde}, {Hristov}, {Huang}, {Imhof}, {Isaak},
  {Israelsson}, {Ivison}, {Jennings}, {Kiernan}, {King}, {Lange}, {Latter},
  {Laurent}, {Laurent}, {Leeks}, {Lellouch}, {Levenson}, {Li}, {Li},
  {Lilienthal}, {Lim}, {Liu}, {Lu}, {Madden}, {Mainetti}, {Marliani}, {McKay},
  {Mercier}, {Molinari}, {Morris}, {Moseley}, {Mulder}, {Mur}, {Naylor},
  {Nguyen}, {O'Halloran}, {Oliver}, {Olofsson}, {Olofsson}, {Orfei}, {Page},
  {Pain}, {Panuzzo}, {Papageorgiou}, {Parks}, {Parr-Burman}, {Pearce},
  {Pearson}, {P{\'e}rez-Fournon}, {Pinsard}, {Pisano}, {Podosek}, {Pohlen},
  {Polehampton}, {Pouliquen}, {Rigopoulou}, {Rizzo}, {Roseboom}, {Roussel},
  {Rowan-Robinson}, {Rownd}, {Saraceno}, {Sauvage}, {Savage}, {Savini},
  {Sawyer}, {Scharmberg}, {Schmitt}, {Schneider}, {Schulz}, {Schwartz},
  {Shafer}, {Shupe}, {Sibthorpe}, {Sidher}, {Smith}, {Smith}, {Smith},
  {Spencer}, {Stobie}, {Sudiwala}, {Sukhatme}, {Surace}, {Stevens}, {Swinyard},
  {Trichas}, {Tourette}, {Triou}, {Tseng}, {Tucker}, {Turner}, {Vaccari},
  {Valtchanov}, {Vigroux}, {Virique}, {Voellmer}, {Walker}, {Ward}, {Waskett},
  {Weilert}, {Wesson}, {White}, {Whitehouse}, {Wilson}, {Winter}, {Woodcraft},
  {Wright}, {Xu}, {Zavagno}, {Zemcov}, {Zhang}, \& {Zonca}}]{griffin10}
{Griffin}, M.~J., {Abergel}, A., {Abreu}, A., {et~al.} 2010, \aap, 518, L3

\bibitem[{{Hatch} {et~al.}(2005){Hatch}, {Crawford}, {Fabian}, \&
  {Johnstone}}]{hatch05}
{Hatch}, N.~A., {Crawford}, C.~S., {Fabian}, A.~C., \& {Johnstone}, R.~M. 2005,
  \mnras, 358, 765

\bibitem[{{Herrera-Camus} {et~al.}(2014){Herrera-Camus}, {Bolatto}, {Wolfire},
  {Smith}, {Croxall}, {Kennicutt}, {Calzetti}, {Helou}, {Walter}, {Leroy},
  {Draine}, {Brandl}, {Armus}, {Sandstrom}, {Dale}, {Aniano}, {Meidt},
  {Boquien}, {Hunt}, {Galametz}, {Tabatabaei}, {Murphy}, {Appleton}, {Roussel},
  {Engelbracht}, \& {Beirao}}]{herrera-camus14}
{Herrera-Camus}, R., {Bolatto}, A.~D., {Wolfire}, M.~G., {et~al.} 2014, ArXiv
  e-prints, arXiv:1409.7123

\bibitem[{{Hicks} \& {Mushotzky}(2005)}]{hicks05}
{Hicks}, A.~K., \& {Mushotzky}, R. 2005, \apjl, 635, L9

\bibitem[{{Jaffe} \& {Bremer}(1997)}]{jaffe97}
{Jaffe}, W., \& {Bremer}, M.~N. 1997, \mnras, 284, L1

\bibitem[{{Jaffe} {et~al.}(2005){Jaffe}, {Bremer}, \& {Baker}}]{jaffe05}
{Jaffe}, W., {Bremer}, M.~N., \& {Baker}, K. 2005, \mnras, 360, 748

\bibitem[{{Johnstone} {et~al.}(2012){Johnstone}, {Canning}, {Fabian},
  {Ferland}, {Lykins}, {Porter}, {van Hoof}, \& {Williams}}]{johnstone12}
{Johnstone}, R.~M., {Canning}, R.~E.~A., {Fabian}, A.~C., {et~al.} 2012,
  \mnras, 425, 1421

\bibitem[{{Johnstone} {et~al.}(1987){Johnstone}, {Fabian}, \&
  {Nulsen}}]{johnstone87}
{Johnstone}, R.~M., {Fabian}, A.~C., \& {Nulsen}, P.~E.~J. 1987, \mnras, 224,
  75

\bibitem[{{Johnstone} {et~al.}(2007){Johnstone}, {Hatch}, {Ferland}, {Fabian},
  {Crawford}, \& {Wilman}}]{johnstone07}
{Johnstone}, R.~M., {Hatch}, N.~A., {Ferland}, G.~J., {et~al.} 2007, \mnras,
  382, 1246

\bibitem[{{Kaastra} {et~al.}(2001){Kaastra}, {Ferrigno}, {Tamura}, {Paerels},
  {Peterson}, \& {Mittaz}}]{kaastra2001}
{Kaastra}, J.~S., {Ferrigno}, C., {Tamura}, T., {et~al.} 2001, \aap, 365, L99

\bibitem[{{Kennicutt}(1998)}]{kennicutt98}
{Kennicutt}, Jr., R.~C. 1998, \araa, 36, 189

\bibitem[{{Klaas} {et~al.}(2001){Klaas}, {Haas}, {M{\"u}ller}, {Chini},
  {Schulz}, {Coulson}, {Hippelein}, {Wilke}, {Albrecht}, \& {Lemke}}]{klaas01}
{Klaas}, U., {Haas}, M., {M{\"u}ller}, S.~A.~H., {et~al.} 2001, \aap, 379, 823

\bibitem[{{Lim} {et~al.}(2012){Lim}, {Ohyama}, {Chi-Hung}, {Dinh-V-Trung}, \&
  {Shiang-Yu}}]{lim12}
{Lim}, J., {Ohyama}, Y., {Chi-Hung}, Y., {Dinh-V-Trung}, \& {Shiang-Yu}, W.
  2012, \apj, 744, 112

\bibitem[{{McDonald} {et~al.}(2014){McDonald}, {Roediger}, {Veilleux}, \&
  {Ehlert}}]{mcdonald14}
{McDonald}, M., {Roediger}, J., {Veilleux}, S., \& {Ehlert}, S. 2014, \apjl,
  791, L30

\bibitem[{{McDonald} \& {Veilleux}(2009)}]{mcdonald09}
{McDonald}, M., \& {Veilleux}, S. 2009, \apjl, 703, L172

\bibitem[{{McDonald} {et~al.}(2011{\natexlab{a}}){McDonald}, {Veilleux}, \&
  {Mushotzky}}]{mcdonald11a}
{McDonald}, M., {Veilleux}, S., \& {Mushotzky}, R. 2011{\natexlab{a}}, \apj,
  731, 33

\bibitem[{{McDonald} {et~al.}(2012{\natexlab{a}}){McDonald}, {Veilleux}, \&
  {Rupke}}]{mcdonald12a}
{McDonald}, M., {Veilleux}, S., \& {Rupke}, D.~S.~N. 2012{\natexlab{a}}, \apj,
  746, 153

\bibitem[{{McDonald} {et~al.}(2010){McDonald}, {Veilleux}, {Rupke}, \&
  {Mushotzky}}]{mcdonald10}
{McDonald}, M., {Veilleux}, S., {Rupke}, D.~S.~N., \& {Mushotzky}, R. 2010,
  \apj, 721, 1262

\bibitem[{{McDonald} {et~al.}(2011{\natexlab{b}}){McDonald}, {Veilleux},
  {Rupke}, {Mushotzky}, \& {Reynolds}}]{mcdonald11b}
{McDonald}, M., {Veilleux}, S., {Rupke}, D.~S.~N., {Mushotzky}, R., \&
  {Reynolds}, C. 2011{\natexlab{b}}, \apj, 734, 95

\bibitem[{{McDonald} {et~al.}(2012{\natexlab{b}}){McDonald}, {Wei}, \&
  {Veilleux}}]{mcdonald12b}
{McDonald}, M., {Wei}, L.~H., \& {Veilleux}, S. 2012{\natexlab{b}}, \apjl, 755,
  L24

\bibitem[{{McDonald} {et~al.}(2012{\natexlab{c}}){McDonald}, {Bayliss},
  {Benson}, {Foley}, {Ruel}, {Sullivan}, {Veilleux}, {Aird}, {Ashby}, {Bautz},
  {Bazin}, {Bleem}, {Brodwin}, {Carlstrom}, {Chang}, {Cho}, {Clocchiatti},
  {Crawford}, {Crites}, {de Haan}, {Desai}, {Dobbs}, {Dudley}, {Egami},
  {Forman}, {Garmire}, {George}, {Gladders}, {Gonzalez}, {Halverson},
  {Harrington}, {High}, {Holder}, {Holzapfel}, {Hoover}, {Hrubes}, {Jones},
  {Joy}, {Keisler}, {Knox}, {Lee}, {Leitch}, {Liu}, {Lueker}, {Luong-van},
  {Mantz}, {Marrone}, {McMahon}, {Mehl}, {Meyer}, {Miller}, {Mocanu}, {Mohr},
  {Montroy}, {Murray}, {Natoli}, {Padin}, {Plagge}, {Pryke}, {Rawle},
  {Reichardt}, {Rest}, {Rex}, {Ruhl}, {Saliwanchik}, {Saro}, {Sayre},
  {Schaffer}, {Shaw}, {Shirokoff}, {Simcoe}, {Song}, {Spieler}, {Stalder},
  {Staniszewski}, {Stark}, {Story}, {Stubbs}, {{\v S}uhada}, {van Engelen},
  {Vanderlinde}, {Vieira}, {Vikhlinin}, {Williamson}, {Zahn}, \&
  {Zenteno}}]{mcdonald12c}
{McDonald}, M., {Bayliss}, M., {Benson}, B.~A., {et~al.} 2012{\natexlab{c}},
  \nat, 488, 349

\bibitem[{{McNamara} {et~al.}(2009){McNamara}, {Kazemzadeh}, {Rafferty},
  {B{\^i}rzan}, {Nulsen}, {Kirkpatrick}, \& {Wise}}]{mcnamara09}
{McNamara}, B.~R., {Kazemzadeh}, F., {Rafferty}, D.~A., {et~al.} 2009, \apj,
  698, 594

\bibitem[{{McNamara} {et~al.}(2000){McNamara}, {Wise}, {Nulsen}, {David},
  {Sarazin}, {Bautz}, {Markevitch}, {Vikhlinin}, {Forman}, {Jones}, \&
  {Harris}}]{mcnamara00}
{McNamara}, B.~R., {Wise}, M., {Nulsen}, P.~E.~J., {et~al.} 2000, \apjl, 534,
  L135

\bibitem[{{Million} {et~al.}(2010{\natexlab{a}}){Million}, {Allen}, {Werner},
  \& {Taylor}}]{million10a}
{Million}, E.~T., {Allen}, S.~W., {Werner}, N., \& {Taylor}, G.~B.
  2010{\natexlab{a}}, \mnras, 405, 1624

\bibitem[{{Million} {et~al.}(2010{\natexlab{b}}){Million}, {Werner},
  {Simionescu}, {Allen}, {Nulsen}, {Fabian}, {B{\"o}hringer}, \&
  {Sanders}}]{million10b}
{Million}, E.~T., {Werner}, N., {Simionescu}, A., {et~al.} 2010{\natexlab{b}},
  \mnras, 407, 2046

\bibitem[{{Mittal} {et~al.}(2011){Mittal}, {O'Dea}, {Ferland}, {Oonk}, {Edge},
  {Canning}, {Russell}, {Baum}, {B{\"o}hringer}, {Combes}, {Donahue}, {Fabian},
  {Hatch}, {Hoffer}, {Johnstone}, {McNamara}, {Salom{\'e}}, \&
  {Tremblay}}]{mittal11}
{Mittal}, R., {O'Dea}, C.~P., {Ferland}, G., {et~al.} 2011, \mnras, 418, 2386

\bibitem[{{Mittal} {et~al.}(2012){Mittal}, {Oonk}, {Ferland}, {Edge}, {O'Dea},
  {Baum}, {Whelan}, {Johnstone}, {Combes}, {Salom{\'e}}, {Fabian}, {Tremblay},
  {Donahue}, \& {Russell}}]{mittal12}
{Mittal}, R., {Oonk}, J.~B.~R., {Ferland}, G.~J., {et~al.} 2012, \mnras, 426,
  2957

\bibitem[{{M{\"u}ller} {et~al.}(2014){M{\"u}ller}, {Balog}, {Nielbock}, {Lim},
  {Teyssier}, {Olberg}, {Klaas}, {Linz}, {Altieri}, {Pearson}, {Bendo}, \&
  {Vilenius}}]{muller14}
{M{\"u}ller}, T., {Balog}, Z., {Nielbock}, M., {et~al.} 2014, Experimental
  Astronomy, 37, 253

\bibitem[{{O'Dea} {et~al.}(2008){O'Dea}, {Baum}, {Privon}, {Noel-Storr},
  {Quillen}, {Zufelt}, {Park}, {Edge}, {Russell}, {Fabian}, {Donahue},
  {Sarazin}, {McNamara}, {Bregman}, \& {Egami}}]{odea08}
{O'Dea}, C.~P., {Baum}, S.~A., {Privon}, G., {et~al.} 2008, \apj, 681, 1035

\bibitem[{{O'Dea} {et~al.}(2010){O'Dea}, {Quillen}, {O'Dea}, {Tremblay},
  {Snios}, {Baum}, {Christiansen}, {Noel-Storr}, {Edge}, {Donahue}, \&
  {Voit}}]{odea10}
{O'Dea}, K.~P., {Quillen}, A.~C., {O'Dea}, C.~P., {et~al.} 2010, \apj, 719,
  1619

\bibitem[{{Oonk} {et~al.}(2010){Oonk}, {Jaffe}, {Bremer}, \& {van
  Weeren}}]{oonk10}
{Oonk}, J.~B.~R., {Jaffe}, W., {Bremer}, M.~N., \& {van Weeren}, R.~J. 2010,
  \mnras, 405, 898

\bibitem[{{Poglitsch} {et~al.}(2010){Poglitsch}, {Waelkens}, {Geis},
  {Feuchtgruber}, {Vandenbussche}, {Rodriguez}, {Krause}, {Renotte}, {van
  Hoof}, {Saraceno}, {Cepa}, {Kerschbaum}, {Agn{\`e}se}, {Ali}, {Altieri},
  {Andreani}, {Augueres}, {Balog}, {Barl}, {Bauer}, {Belbachir}, {Benedettini},
  {Billot}, {Boulade}, {Bischof}, {Blommaert}, {Callut}, {Cara}, {Cerulli},
  {Cesarsky}, {Contursi}, {Creten}, {De Meester}, {Doublier}, {Doumayrou},
  {Duband}, {Exter}, {Genzel}, {Gillis}, {Gr{\"o}zinger}, {Henning},
  {Herreros}, {Huygen}, {Inguscio}, {Jakob}, {Jamar}, {Jean}, {de Jong},
  {Katterloher}, {Kiss}, {Klaas}, {Lemke}, {Lutz}, {Madden}, {Marquet},
  {Martignac}, {Mazy}, {Merken}, {Montfort}, {Morbidelli}, {M{\"u}ller},
  {Nielbock}, {Okumura}, {Orfei}, {Ottensamer}, {Pezzuto}, {Popesso},
  {Putzeys}, {Regibo}, {Reveret}, {Royer}, {Sauvage}, {Schreiber}, {Stegmaier},
  {Schmitt}, {Schubert}, {Sturm}, {Thiel}, {Tofani}, {Vavrek}, {Wetzstein},
  {Wieprecht}, \& {Wiezorrek}}]{poglitsch10}
{Poglitsch}, A., {Waelkens}, C., {Geis}, N., {et~al.} 2010, \aap, 518, L2

\bibitem[{{Roediger} {et~al.}(2014){Roediger}, {Br{\"u}ggen}, {Owers},
  {Ebeling}, \& {Sun}}]{roediger14}
{Roediger}, E., {Br{\"u}ggen}, M., {Owers}, M.~S., {Ebeling}, H., \& {Sun}, M.
  2014, \mnras, 443, L114

\bibitem[{{Rosa-Gonz{\'a}lez} {et~al.}(2002){Rosa-Gonz{\'a}lez}, {Terlevich},
  \& {Terlevich}}]{rg02}
{Rosa-Gonz{\'a}lez}, D., {Terlevich}, E., \& {Terlevich}, R. 2002, \mnras, 332,
  283

\bibitem[{{Salom{\'e}} \& {Combes}(2003)}]{salome03}
{Salom{\'e}}, P., \& {Combes}, F. 2003, \aap, 412, 657

\bibitem[{{Sharma} {et~al.}(2012){Sharma}, {McCourt}, {Quataert}, \&
  {Parrish}}]{sharma12}
{Sharma}, P., {McCourt}, M., {Quataert}, E., \& {Parrish}, I.~J. 2012, \mnras,
  420, 3174

\bibitem[{{Sun}(2009)}]{sun09b}
{Sun}, M. 2009, \apj, 704, 1586

\bibitem[{{Sun} {et~al.}(2009){Sun}, {Voit}, {Donahue}, {Jones}, {Forman}, \&
  {Vikhlinin}}]{sun09a}
{Sun}, M., {Voit}, G.~M., {Donahue}, M., {et~al.} 2009, \apj, 693, 1142

\bibitem[{{Veilleux} {et~al.}(2010){Veilleux}, {Weiner}, {Rupke}, {McDonald},
  {Birk}, {Bland-Hawthorn}, {Dressler}, {Hare}, {Osip}, {Pietraszewski}, \&
  {Vogel}}]{mmtf}
{Veilleux}, S., {Weiner}, B.~J., {Rupke}, D.~S.~N., {et~al.} 2010, \aj, 139,
  145

\bibitem[{{Voit} {et~al.}(2008){Voit}, {Cavagnolo}, {Donahue}, {Rafferty},
  {McNamara}, \& {Nulsen}}]{voit08}
{Voit}, G.~M., {Cavagnolo}, K.~W., {Donahue}, M., {et~al.} 2008, \apjl, 681, L5

\bibitem[{{Weingartner} \& {Draine}(2001)}]{weingartner01}
{Weingartner}, J.~C., \& {Draine}, B.~T. 2001, \apj, 548, 296

\bibitem[{{Werner} {et~al.}(2007){Werner}, {Kaastra}, {Takei}, {Lieu}, {Vink},
  \& {Tamura}}]{werner07}
{Werner}, N., {Kaastra}, J.~S., {Takei}, Y., {et~al.} 2007, \aap, 468, 849

\bibitem[{{Werner} {et~al.}(2011){Werner}, {Sun}, {Bagchi}, {Allen}, {Taylor},
  {Sirothia}, {Simionescu}, {Million}, {Jacob}, \& {Donahue}}]{werner11}
{Werner}, N., {Sun}, M., {Bagchi}, J., {et~al.} 2011, \mnras, 415, 3369

\bibitem[{{Werner} {et~al.}(2014){Werner}, {Oonk}, {Sun}, {Nulsen}, {Allen},
  {Canning}, {Simionescu}, {Hoffer}, {Connor}, {Donahue}, {Edge}, {Fabian},
  {von der Linden}, {Reynolds}, \& {Ruszkowski}}]{werner14}
{Werner}, N., {Oonk}, J.~B.~R., {Sun}, M., {et~al.} 2014, \mnras, 439, 2291

\bibitem[{{ZuHone} {et~al.}(2010){ZuHone}, {Markevitch}, \&
  {Johnson}}]{zuhone10}
{ZuHone}, J.~A., {Markevitch}, M., \& {Johnson}, R.~E. 2010, \apj, 717, 908

\bibitem[{{ZuHone} {et~al.}(2011){ZuHone}, {Markevitch}, \& {Lee}}]{zuhone11}
{ZuHone}, J.~A., {Markevitch}, M., \& {Lee}, D. 2011, \apj, 743, 16

\end{thebibliography}

% ===== APPENDIX ====== %
\end{document}